\documentclass[sigplan,10pt]{acmart}

\AtBeginDocument{%
  }

\copyrightyear{2026}
\acmYear{2026}
\setcopyright{cc}
\setcctype{by}
\acmConference[EUROSYS '26]{21st European Conference on Computer Systems}{April 27--30, 2026}{Edinburgh, Scotland Uk}
\acmBooktitle{21st European Conference on Computer Systems (EUROSYS '26), April 27--30, 2026, Edinburgh, Scotland Uk}
\acmPrice{}
\acmDOI{10.1145/3767295.3769378}
\acmISBN{979-8-4007-2212-7/26/04}

\usepackage{algorithm}
\usepackage{algpseudocode}
\usepackage{soul}
\usepackage{float}
\usepackage{graphicx}
\usepackage{subcaption}
\usepackage[english]{babel}
\usepackage{makecell}
\usepackage{pbalance}

\begin{document}

\title{Matrix‑PIC: Harnessing Matrix Outer-product for High‑Performance Particle‑in‑Cell Simulations}


\author{Yizhuo Rao}
\orcid{0009-0000-3572-6969}
\affiliation{%
    \institution{Sun Yat-Sen University}
    \city{Guangzhou}
    \country{China}
}
\email{raoyzh6@mail2.sysu.edu.cn}

\author{Xingjian Cui}
\affiliation{%
    \institution{Sun Yat-Sen University}
    \city{Guangzhou}
    \country{China}
}
\email{cuixj8@mail2.sysu.edu.cn}

\author{Jiabin Xie}
\affiliation{%
    \institution{Sun Yat-Sen University}
    \city{Guangzhou}
    \country{China}
}
\email{xiejb6@mail2.sysu.edu.cn}

\author{Shangzhi Pang}
\affiliation{%
    \institution{Sun Yat-Sen University}
    \city{Guangzhou}
    \country{China}
}
\email{pangshzh@mail2.sysu.edu.cn}

\author{Guangnan Feng}
\affiliation{%
    \institution{Sun Yat-Sen University}
    \city{Guangzhou}
    \country{China}
}
\email{fenggn7@mail.sysu.edu.cn}
\authornote{Corresponding authors.}

\author{Jinhui Wei}
\affiliation{%
    \institution{Sun Yat-Sen University}
    \city{Guangzhou}
    \country{China}
}
\email{weijh28@mail2.sysu.edu.cn}

\author{Zhiguang Chen}
\affiliation{%
    \institution{Sun Yat-Sen University}
    \city{Guangzhou}
    \country{China}
}
\email{zhiguang.chen@nscc-gz.cn}
% \authornote{Corresponding authors.}
\authornotemark[1] 

\author{Yutong Lu}
\affiliation{%
    \institution{Sun Yat-Sen University}
    \city{Guangzhou}
    \country{China}
}
\email{luyutong@mail.sysu.edu.cn}
% \affiliation{%
%   \institution{National Supercomputer Center in Guangzhou}
%   \city{Guangzhou}
%   \country{China}
% }
% \affiliation{%
%   \institution{National Supercomputer Center in Shenzhen}
%   \city{Shenzhen}
%   \country{China}
% }

\renewcommand{\shortauthors}{Yizhuo Rao et al.}

\begin{abstract}

Particle-in-Cell (PIC) simulations devote most cycles to particle-grid interactions, and their fine-grained atomic updates become a severe bottleneck on traditional many-core CPUs. The evolution of CPU architectures, particularly the integration of specialized Matrix Processing Units (MPUs) designed for efficient matrix outer-product operations, presents a paradigm shift and an opportunity to alleviate these bottlenecks. Capitalizing on this architectural advancement, this work focuses on adapting the critical current deposition step in PIC simulations to this new matrix-centric computational model.

We introduce \textbf{MatrixPIC}, a novel framework representing, to our knowledge, the first holistic co-design of the deposition kernel, data layout, and an incremental sorting mechanism, all tailored specifically for the hybrid MPU-VPU SIMD execution model on modern CPUs. It provides a validated design paradigm for the scientific computing community to leverage the next generation of high-density hardware. The key technical innovations are: (i) the refactoring of the current deposition algorithm into block-matrix updates that map efficiently onto the MPU's native operational paradigm; (ii) a hybrid execution pipeline that synergistically orchestrates MPU kernels for high-density accumulation with VPU stages for data preparation and control flow; and (iii) an $O(1)$-amortized incremental particle sorter, which utilizes a gapped packed-memory array to establish and maintain the crucial data locality required for MPU efficiency.

Evaluated on a next-generation high-performance computing platform, \textbf{MatrixPIC} demonstrates substantial performance improvements. In complex Laser-Wakefield Acceleration (LWFA) simulations, it yields up to a \textbf{2.63$\times$} speedup in total simulation time. Most significantly, for a higher-order, third-order deposition scheme, \textbf{MatrixPIC} accelerates the core deposition kernel by \textbf{8.7$\times$} over the original baseline and is \textbf{2.0$\times$} faster than our best hand-tuned VPU implementation. Furthermore, our CPU-based framework reaches \textbf{83.08\%} of theoretical peak performance, a hardware utilization rate nearly \textbf{2.8$\times$} higher than a highly-optimized CUDA kernel on a data center GPU. These results underscore the transformative potential of co-designing scientific algorithms for emerging, matrix-capable CPU architectures, paving the way for enhanced performance in large-scale computations.

\end{abstract}



% \keywords{Do, Not, Us, This, Code, Put, the, Correct, Terms, for,


\maketitle
\section{Introduction}
\label{sec:intro}

Particle-based mesh algorithms represent a cornerstone of simulation across diverse scientific domains. These include Particle-in-Cell (PIC) method in plasma physics~\cite{SMILEI2018, WarpX2022}, the Particle-Mesh (PM) method in astrophysical N-body simulations ~\cite{couchman1991mesh, breton2025pysco}, and the Particle-Mesh-Ewald (PME) method in molecular dynamics \cite{cheatham1995molecular, abraham2011optimization}. Within plasma physics, PIC simulation is particularly vital for studying complex kinetic phenomena, from laser-plasma interactions \cite{SMILEI2018, WarpX2022} to astrophysical plasmas \cite{SMILEI2018}.

\begin{figure}[t]
  \centering
  \includegraphics[width=\linewidth]{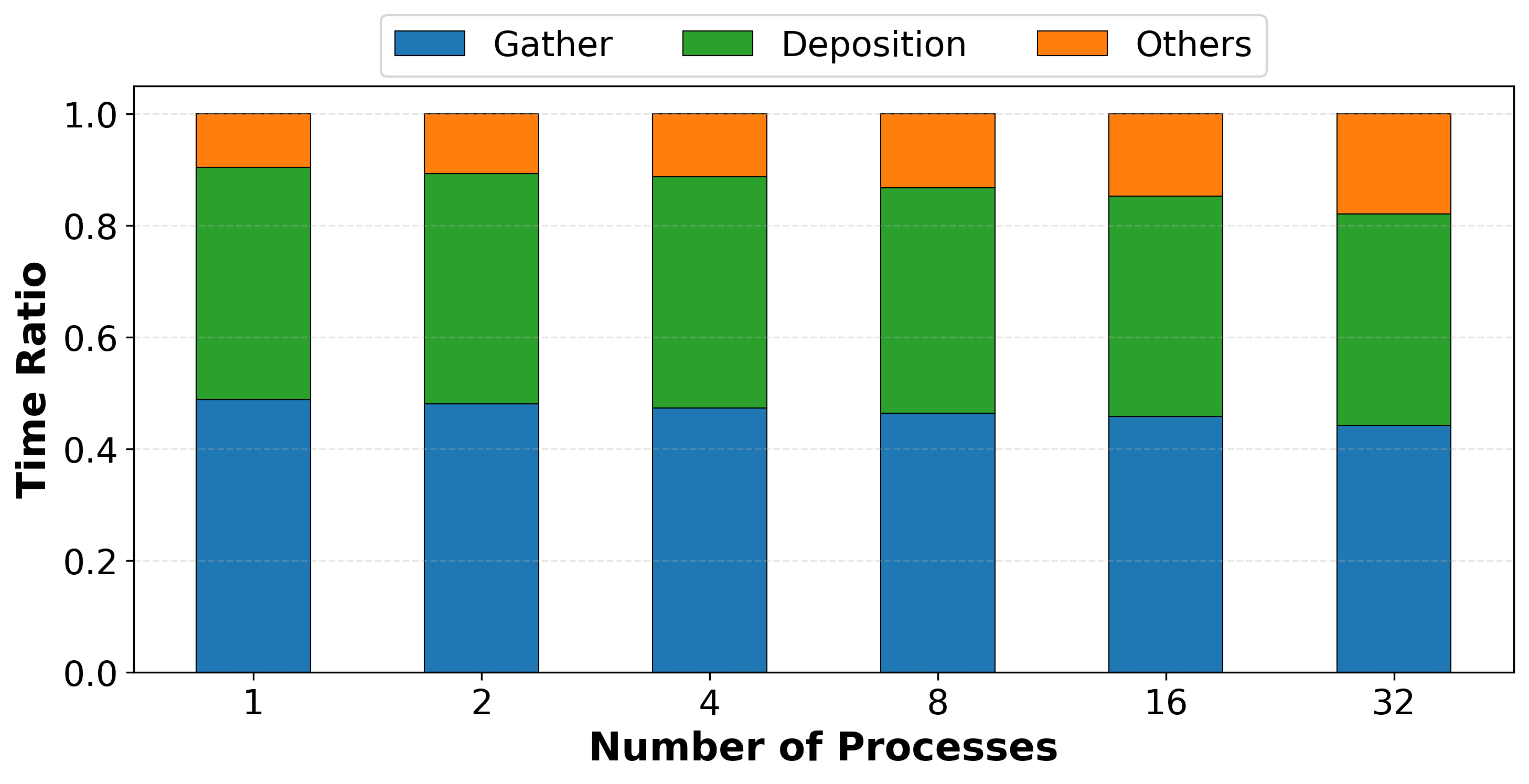}
  \caption{Runtime breakdown of uniform plasma PIC simulation on the Tianhe Xingyi HPC platform. (We utilize \texttt{WarpX v24.07}\cite{WarpX2022} with a ~30-million cells and ~4.3 billion particles)}
  \label{fig:compute_cost}
  \Description{A pie chart or bar graph showing that current deposition takes a large percentage of PIC simulation time.}
\end{figure}

A critical computational kernel within these particle-mesh simulations is the \emph{deposition} step, which often constitutes a significant performance bottleneck, accounting for 40-70\% of the total execution time on many-core CPUs \cite{Barsamian2018StrictBinning, Nakashima2017PIC}.
For instance, our scalability tests on 32 processes reveal a significant performance bottleneck: the particle deposition and gather steps combined account for over 80\% of the total execution time. As illustrated in Figure~\ref{fig:compute_cost}, the deposition step by itself constitutes more than 40\% of this total.
% For instance, our scalability tests on the Tianhe Xingyi platform demonstrate that with 32 processes, the particle deposition and gather steps together account for more than 80\% of the execution time, with \emph{deposition} alone exceeding 40\%, as illustrated in Figure~\ref{fig:compute_cost}.

At its core, this performance bottleneck stems from a fundamental computational pattern: \textit{accumulating a large number of sparse, localized updates onto a regular, dense grid}. This pattern can be deconstructed into three key elements: a \textbf{Source} of discrete entities carrying physical quantities; a \textbf{Target} representing a regular background grid; and an \textbf{Operation}, which is a "scatter-add" or "deposition" process.

In the context of the \textbf{PIC method} in plasma simulations~\cite{SMILEI2018, WarpX2022}, this pattern manifests with a source of charged particles, a 3D grid as the target, and an operation that uses a shape function (e.g., Cloud-in-Cell or CIC) to scatter each particle's charge for solving Maxwell's equations. This process is algorithmically isomorphic to the mass deposition step in the \textbf{Particle-Mesh (PM) method} used in N-body cosmology simulations~\cite{couchman1991mesh, breton2025pysco}. In that domain, the source consists of galaxies or dark matter particles, the target is a grid spanning cosmic space, and the operation uses a shape function to deposit mass for solving the gravitational potential on the grid. These isomorphic patterns are detailed further in Appendix~\ref{appen:generalizing}.

The computational challenges of this deposition pattern are twofold: poor data locality stemming from the unordered nature of particles, and severe write contention when multiple particles attempt to update the same grid node. These issues lead to a high cache miss rate and significant inter-thread write conflicts \cite{Barsamian2018StrictBinning, Nakashima2017PIC}.

Overcoming this memory-bound challenge through Single-Instruction-Multiple-Data (SIMD) vectorization has been a primary focus of optimization efforts \cite{Beck2019AdaptiveSIMD, Vincenti2017SIMD}. However, SIMD-based deposition is challenged by atomic write conflicts that arise when multiple SIMD lanes concurrently update the same grid point. Existing solutions mitigate these conflicts using temporary cell-local buffers (e.g., the \texttt{jtmp} approach) \cite{Beck2019AdaptiveSIMD} or novel data structures like \texttt{rhocell} \cite{Vincenti2017SIMD}. For instance, the \texttt{rhocell} data structure eliminates intra-vector write conflicts by reframing the SIMD operation, enabling a single particle's contribution to update its eight neighboring grid points without requiring an \texttt{atomicADD} \cite{Vincenti2017SIMD}.

% While effective, these strategies are often constrained by the bandwidth and computational width of conventional Vector Processing Units (VPUs). The ongoing evolution of hardware architectures presents new opportunities. Modern processors, such as Apple's M4 \cite{bib_apple_com, Remke2024HelloSME} and emerging deeply integrated architectures like the NVIDIA Grace Hopper \cite{simakov2024first} and AMD MI300 series \cite{bib_AMD_MI300X}, now integrate Matrix Processing Units (MPUs) for vector outer product operations. These MPUs \cite{Zhao2024StencilSME, Remke2024HelloSME} offer significantly higher computational density than conventional VPUs, presenting a promising avenue for accelerating hotspot kernels like current deposition. However, effectively harnessing these MPUs for such simulations introduces distinct challenges:
% \begin{enumerate}
%     \item \textbf{Algorithmic Adaptation:} Adapting algorithms from the sparse particle-grid interactions of PIC to the dense matrix operations favored by MPUs.
%     \item \textbf{Hybrid Computation Model:} Coordinating MPUs with general-purpose VPUs, which are required for unsupported operations like conditional branching and complex data manipulation.
%     \item \textbf{Data Locality and Register Management:} Managing register-level data locality and minimizing data movement between these hybrid processing units to exploit their full potential.
% \end{enumerate}

While effective, these strategies are often constrained by the bandwidth and computational width of conventional Vector Processing Units (VPUs). The ongoing evolution of hardware architectures presents new opportunities. Modern processors, such as Apple's M4 \cite{bib_apple_com, Remke2024HelloSME} and emerging deeply integrated architectures like the NVIDIA Grace Hopper \cite{simakov2024first} and AMD MI300 series \cite{bib_AMD_MI300X}, now integrate Matrix Processing Units (MPUs) for vector outer product operations. These MPUs \cite{Zhao2024StencilSME, Remke2024HelloSME} offer significantly higher computational density than conventional VPUs, presenting a promising avenue for accelerating hotspot kernels like current deposition. 

However, effectively harnessing these MPUs for such simulations introduces distinct challenges. \textbf{First}, algorithmic adaptation is required to transition algorithms from the sparse particle-grid interactions of PIC to the dense matrix operations favored by MPUs. \textbf{Second}, a \textit{hybrid computation model} is required to coordinate MPUs with general-purpose VPUs, which are necessary for unsupported operations like conditional branching. \textbf{Finally}, success hinges on effective \textit{data locality and register management} to minimize data movement between these hybrid processing units and exploit their full potential.

This paper addresses these challenges by proposing \textbf{MatrixPIC}, a novel algorithmic framework for current deposition in particle-mesh codes. To our knowledge, \textbf{MatrixPIC} is the first holistic co-design that maps the PIC deposition kernel to CPUs equipped with MPUs, providing a validated design paradigm for how the scientific computing community can embrace the next generation of high-density computing hardware. Our key contributions are:
% \vspace{-6pt}
\begin{itemize}
\item A matrix outer product-based computation model that maps particle deposition to the MPU paradigm, inherently resolving atomic write conflicts.
\item A hybrid MPU-VPU computation kernel that synergistically combines high-density MPU accumulation with VPU-based data preparation and control flow.
\item A fine-grained, low-overhead incremental particle sorting algorithm with O(1) amortized cost, designed to establish and maintain the data locality crucial for MPU efficiency.
\item A multi-level data reorganization strategy that integrates local sorting with periodic global resorting to preserve an efficient Structure-of-Arrays (SoA) data layout.
\end{itemize}

We implement and evaluate our proposed methods within the open-source PIC code WarpX \cite{WarpX2022}, a highly optimized framework renowned in the plasma physics community and a recipient of the ACM Gordon Bell Prize. Our experiments, conducted on a domestic High-Performance Computing (HPC) platform equipped with an LX2 CPU, demonstrate significant and broad performance improvements. For standard first-order (CIC) schemes, \textbf{MatrixPIC} achieves up to a \textbf{1.19$\times$} speedup in total simulation time for uniform plasma workloads and a more substantial \textbf{2.63$\times$} speedup in complex laser-wakefield acceleration scenarios. The benefits of our MPU-centric co-design are most pronounced with computationally intensive, higher-order algorithms; for the third-order (QSP) deposition scheme, our framework accelerates the core kernel by a remarkable \textbf{8.7$\times$} over the baseline in uniform plasma workloads. Furthermore, a cross-platform efficiency analysis reveals that our approach achieves a hardware utilization nearly \textbf{2.8$\times$} than the highly-optimized WarpX CUDA implementation on a data center GPU, underscoring the profound effectiveness of our co-design paradigm.

\section{Related Work}
\label{sec:related_work}
The performance of Particle-in-Cell (PIC) simulations, particularly the current deposition step, has been widely studied due to its pivotal role in modeling plasma phenomena on modern computing platforms~\citep{SMILEI2018, WarpX2022, Li2021YHPIC}. Research spans algorithmic enhancements~\citep{Nakashima2017PIC, Beck2019AdaptiveSIMD}, data structure improvements~\cite{Vincenti2017SIMD,Barsamian2017DataStructures,Barsamian2018DataLayouts}, and adaptations to evolving hardware~\citep{WarpX2022,Bird2022VPIC2,Bowers2008VPIC}, including parallelization and vectorization strategies~\citep{Barsamian2017DataStructures, Barsamian2018DataLayouts}. This section reviews prior work relevant to our approach, categorized by optimization focus.

\subsection{SIMD Optimization for Current Deposition}
Current deposition in PIC simulations is computationally demanding due to irregular memory access and write conflicts when updating grid quantities, making it a key target for SIMD vectorization on multi-core and many-core processors~\citep{Nakashima2017PIC, Barsamian2018StrictBinning}. Early efforts tackled atomic update conflicts when multiple particles in a SIMD instruction wrote to overlapping grid points. A prevalent solution uses thread-local buffers (e.g., temporary local buffer $J_{local}$) to accumulate contributions without contention, followed by a global reduction~\citep{Beck2019AdaptiveSIMD}. While this avoids atomics in the particle loop, it incurs memory overhead and can bottleneck at reduction, especially with high thread counts.

A breakthrough came with the \texttt{rhocell} data structure by~\citet{Vincenti2017SIMD}, shifting to a particle-centric approach where each particle’s contribution is SIMD-distributed to neighboring grid cells, eliminating intra-SIMD conflicts. This yielded 2$\times$ to 2.5$\times$ speed-ups on Haswell Xeons with AVX2-256 bits wide data registers. However, the final reduction involves non-contiguous memory accesses, and the method targets traditional VPUs. Our work extends \texttt{rhocell} to exploit the computational density of MPUs~\citep{Zhao2024StencilSME}, introducing a novel direction for PIC deposition.

\subsection{Particle Sorting and Data Locality}
Data locality is critical for PIC efficiency, affecting cache usage, memory bandwidth, and SIMD performance~\citep{Barsamian2017DataStructures, Barsamian2018DataLayouts}. Particle sorting (binning) groups particles by grid cell to enhance locality~\citep{Nakashima2017PIC, Barsamian2018StrictBinning, Beck2019AdaptiveSIMD, Decyk2014PICEmergingArch}.~\citet{Nakashima2017PIC} optimized binning for Intel Xeon Phi, using strict binning with SOA layouts to scalarize field data, while~\citet{Barsamian2018StrictBinning} employed fixed-capacity linked lists for dynamic bins on Skylake. Space-filling curves (e.g., Morton ordering) further improve cache performance~\citep{Barsamian2017DataStructures, SMILEI2018}.

Sorting algorithms like radix and counting sorts are common for initial or infrequent resorts~\citep{Bowers2001HybridSort}, but full sorts per timestep are costly for large, mobile particle sets~\citep{Nakashima2017PIC}. This motivates our lightweight, incremental sorting, inspired by adaptive SIMD techniques that toggle kernels based on particle density~\citep{Beck2019AdaptiveSIMD}, effective with as few as eight particles per cell.

\subsection{PIC Optimizations for Emerging Architectures}

The evolution of massively parallel systems, including GPUs and Many Integrated Core (MIC) architectures, has spurred extensive optimization of Particle-in-Cell (PIC) codes \cite{Decyk2014PICEmergingArch, Li2021YHPIC, WarpX2022}. Leading codes like WarpX \cite{WarpX2022}, VPIC \cite{Bowers2008VPIC, Bird2022VPIC2}, PIConGPU \cite{Bussmann2013PIConGPU, Zenker2016PIConGPUOpenPower, Chandrasekaran2019PIConGPUSummit}, and YHPIC \cite{Li2021YHPIC} now scale to trillions of particles by leveraging domain decomposition and fine-grained kernel tuning.Other efforts have focused on specific architectures, such as SMILEI, adapting codes to exploit features ranging from Arm's Scalable Vector Extension (SVE) to hybrid MPI-OpenMP models \cite{HarrisonA64FXSVE, SMILEI2018}. 

While GPUs are a dominant platform, their specialized matrix hardware (e.g., Tensor Cores), designed for Matrix-Matrix-Accumulate (MMA) operations, is a poor architectural match for the \textit{scatter-add} pattern of PIC deposition. Consequently, leading GPU-PIC codes bypass this hardware and rely on general-purpose CUDA cores \cite{Bussmann2013PIConGPU}. This architectural mismatch highlights an opportunity presented by a new class of CPU-integrated accelerators: Matrix Processing Units (MPUs). These MPUs, based on a Matrix-Outer-Product-Accumulate (MOPA) paradigm, offer a different computational model whose applicability to sparse, particle-based methods remained unexplored.

This work bridges that gap. Building upon recent successes applying MPUs to other scientific kernels like matrix multiplication and stencil computations \cite{Remke2024HelloSME, Zhao2024StencilSME}, we demonstrate for the first time that the PIC deposition pattern can be fundamentally reformulated through algorithmic and data layout co-design to map efficiently onto the MPU's MOPA paradigm. Our approach opens a new optimization pathway, distinct from both traditional VPU vectorization and prevailing GPU-based methods, by harnessing the computational power of these emerging general-purpose matrix accelerators.

\subsection{Dynamic Data Structures for Particle Management}
Managing dynamic sorted datasets is vital for PIC particle sorting. Packed Memory Arrays (PMAs) maintain sorted order with gaps for efficient updates~\citep{Itai1981SparseTables, Bender2000CacheObliviousBtrees}, enhanced by adaptive PMAs for varied insertion patterns~\citep{Bender2007AdaptivePMA}. Our Gapped PMA (GPMA) approach builds on this, using block gaps for O(1) updates~\citep{Durand2012PMA_MovingParticles, Nakashima2017PIC}, ideal for incremental re-sorting under CFL-constrained particle movement~\cite{bib_CFL_Condi_placeholder, WarpX2022}. Unlike full sorts, GPMA manipulates indices, deferring data movement until necessary, complementing locality-focused efforts~\citep{WarpXOptaneIPDPSW2021}.

\begin{figure}[t]
  \centering
  \includegraphics[width=0.8\linewidth]{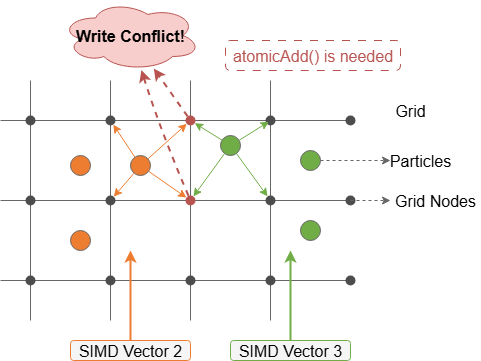}
  \caption{Illustration of SIMD current deposition leading to potential write conflicts on a grid node.}
  \label{fig:simd_atomic_conflict}
\end{figure}

\section{Preliminaries}
\label{sec:prelim}

This section summarizes the background on PIC current deposition, challenges of SIMD vectorization for this step and highlight novel architectural features and data structures that form the basis of our proposed optimizations.

\subsection{PIC Current Deposition}
As established in Section~\ref{sec:intro}, the PIC method models a system where charged particles move continuously through a simulation space while electromagnetic fields are defined on a discrete grid~\cite{Birdsall1985PlasmaPhysics, Hockney1988ComputerSim}. A standard PIC loop iteratively performs several key stages: field interpolation from grid to particles, the particle push according to the Lorentz force, current (or charge) deposition from particles back to the grid, and solving Maxwell's equations on the grid to update fields~\cite{SMILEI2018, WarpX2022}. In this step, each particle $p$ with charge $q_p$ and velocity $\mathbf{v}_p$ contributes to the current density $\mathbf{J}$ at its surrounding grid nodes $\mathbf{x}_i$:
\begin{equation}
\mathbf{J}(\mathbf{x}_i) = \sum_p q_p \mathbf{v}_p S(\mathbf{x}_i - \mathbf{x}_p)
\label{eq:current_deposition_general}
\end{equation}
where \(S\) is the shape function (e.g.\ $S \in \mathbb{R}_{2,3} $ for first-order CIC scheme in 3D)~\cite{Nakashima2017PIC}.  Current deposition is memory‐bound—each particle causes 8 scattered writes—and dominates runtime (40–70\%) on many‐core CPUs~\cite{Barsamian2018StrictBinning, Nakashima2017PIC}. 

\subsection{SIMD and Atomic‐Write Conflicts}
Single Instruction, Multiple Data (SIMD) vectorization, leveraging Vector Processing Units (VPUs), is a key strategy to improve the arithmetic throughput of current deposition by processing multiple particles concurrently~\cite{Beck2019AdaptiveSIMD}. However, a fundamental challenge arises when particles within the same SIMD vector, or processed by different parallel threads, need to update the same grid node. This necessitates the use of atomic operations to ensure correct accumulation and avoid race conditions:
\begin{equation}
\text{GridNode}[i] \leftarrow \texttt{atomicAdd}(J_i,\Delta J_{p_{ip}})
\label{eq:atomic_add}
\end{equation}
These atomic operations can serialize execution and stall SIMD pipelines, particularly in regions with high particle density, thereby limiting overall vectorization efficiency~\cite{Vincenti2017SIMD} (illustrated in Figure~\ref{fig:simd_atomic_conflict}). 

\subsection{Matrix Outer‐Product Units}
Recent CPU architectures are increasingly integrating specialized Matrix Processing Units (MPUs), which natively compute outer products for two vectors $\mathbf{a} \in \mathbb{R}^m$ and $\mathbf{b} \in \mathbb{R}^n$. It can be accumulated into a matrix tile $\mathbf{C} \in \mathbb{R}^{m \times n}$ as:
\begin{equation}
\mathbf{C} \leftarrow \mathbf{C} + \mathbf{a} \otimes \mathbf{b}
\label{eq:outer_product}
\end{equation}
and filling an \(m\times n\) tile per instruction~\cite{Weidman2021SMEArm,Zhao2024StencilSME}.  MPUs deliver much higher FLOP‐to‐byte ratios than VPUs but lack scatter/gather and predicate support, requiring hybrid VPU–MPU pipelines~\cite{Remke2024HelloSME}.  

% \begin{figure}[t]
%   \centering
%   \includegraphics[width=0.8\linewidth]{figs/outer_product_unit.pdf}
%   \caption{Matrix outer‐product: MPU computes \(\mathbf{C}\!+\!=\!\mathbf{a}\otimes\mathbf{b}\) on an \(m\times n\) tile.}
%   \label{fig:outer}
% \end{figure} 

\begin{figure}[t]
  \centering
  % Note: For a side-by-side figure, you might need to adjust the width. 0.8\linewidth is often a good starting point.
  \includegraphics[width=0.8\linewidth]{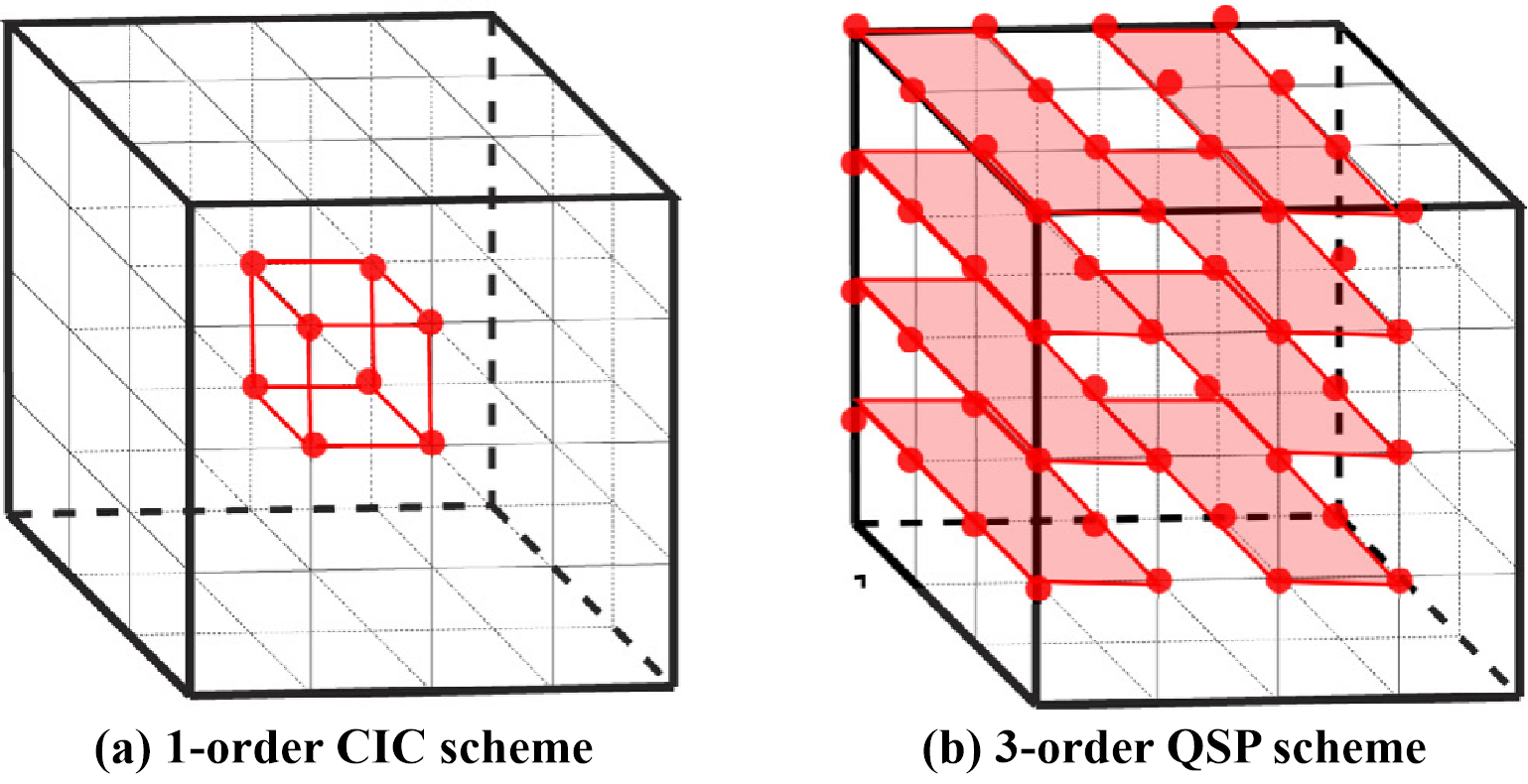} 
  \caption{Conceptual illustration of the \texttt{rhocell} data layout for different interpolation schemes~\cite{Vincenti2017SIMD}. (a) The first-order CIC scheme. (b) The third-order QSP scheme.}
  \label{fig:rhocell_layout}
\end{figure}

\subsection{Rhocell Data Layout}
The \texttt{rhocell} data structure, introduced by ~\citep{Vincenti2017SIMD}, provides an effective mechanism to mitigate atomic write conflicts when using VPU-based SIMD for current deposition. The core idea is to restructure the target of the deposition for each particle. Instead of having particles directly and concurrently updating the global grid array, each particle's contributions to its neighboring grid nodes are first accumulated into a small, temporary, particle-centric array---the \texttt{rhocell}. 

Conceptually, this approach shares its core philosophy with the \texttt{reducer} pattern found in parallel programming models like Cilk's Hyperobjects ~\cite{frigo2009reducers, lee2015efficiently}. Both use private, thread-local storage to mitigate race conditions during parallel updates, deferring a final, conflict-free reduction until all parallel work is complete. The \texttt{rhocell} can thus be viewed as a domain-specific instantiation of this pattern, highly optimized for particle-grid interactions. Our work advances this idea by enhancing rhocell's data locality through sorting and extending its application to the MPU paradigm.

For a CIC scheme in 3D, \texttt{rhocell} typically holds a 2D array
\begin{equation}
\texttt{rhocells}(1:8, i_{cell})\;\in\;\mathbb{R}^{8 \times N_{\mathrm{cells}}},
\label{eq:rhocells}
\end{equation}
where each column \(\texttt{rhocells}(:,i_{cell})\) packs the 8 vertices of cell \(i_{cell}\) contiguously and aligned to a 64‐byte cache line (as shown in Figure~\ref{fig:rhocell_layout}).

% \begin{figure}[h]
%   \centering
%   \includegraphics[width=0.4\linewidth]{figs/rhocell3d.png} % User specified path
%   \caption{Conceptual illustration of the \texttt{rhocell} data layout (CIC). Particle contributions to the 8 neighboring grid nodes are stored contiguously in \texttt{rhocell}.}
%   \label{fig:rhocell_layout}
% \end{figure} 

After all particles have deposited, this method performs one reduction:
\begin{equation}
\rho[i] \;+\!=\; \sum_{i_{cell} \,\ni\, i} \texttt{rhocells}(k(i_{cell},i),\,i_{cell}),
\label{eq:rhocell_reduction}
\end{equation}
where \(k(i_{cell},i)\) maps cell \(i_{cell}\) to its local index \(1\le k\le8\) for vertex \(i\).  
This reduction scans \(N_{\mathrm{cells}}\) entries and is thus \(O(N_{\mathrm{cells}})\), not \(O(N_p)\), making its overhead negligible when \(N_p \!\gg\! N_{\mathrm{cells}}\).  

% While our implementation focuses on first‐order (CIC) deposition, the same principle extends to higher‐order shapes (TSC, QSP) by grouping vertices in planar blocks of eight and applying analogous vector updates~\cite{Vincenti2017SIMD}. 

\subsection{Gapped Packed Memory Array}
\begin{figure}[h]
  \centering
  \includegraphics[width=\linewidth]{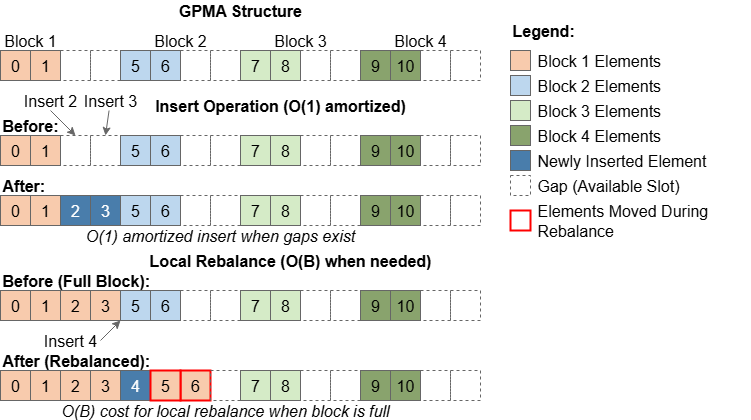}
  \caption{GPMA: gaps between blocks enable efficient local index updates.}
  \label{fig:gpma}
\end{figure}

Frequent particle sorting is necessary to maintain data locality for efficient SIMD processing, especially with MPUs which benefit from highly regular data access. However, full sorting every timestep is computationally expensive. The Gapped Packed Memory Array (GPMA) is a dynamic array structure based on Packed Memory Arrays (PMAs)~\cite{Bender2000CacheObliviousBtrees, Bender2007AdaptivePMA} that supports efficient incremental updates to sorted data. 
The GPMA maintains sorted indices with interspersed gaps, supporting \(O(1)\) amortized insert/delete if slots exist within a block of size \(B\), with rare local rebalances costing \(O(B)\) as shown in Figure~\ref{fig:gpma}. This makes GPMA highly suitable for the incremental sorting strategy employed in our work, where most particles exhibit limited displacement per timestep due to the CFL condition~\cite{bib_CFL_condition_PIC_ref}.

\section{Methodology}
\label{sec:methodology}

\textbf{MatrixPIC} is a framework for optimizing PIC current deposition on modern CPUs equipped with both Vector Processing Units (VPUs) and Matrix Processing Units (MPUs). \textbf{MatrixPIC} integrates MPU-accelerated deposition kernels with a low-overhead incremental particle sorting strategy and an adaptive global resorting policy. This section details the overall algorithmic flow, followed by in-depth descriptions of the MPU-based deposition kernel and the efficient particle sorting mechanism.

\begin{algorithm}
\caption{Overview of the \textbf{MatrixPIC} Current Deposition and Sorting Workflow}
\label{alg:main_pic_loop_overview}
\begin{algorithmic}[1]
  % Initialization phase
  \State \textbf{Initialization (once per MPI rank):}
  \State \quad Read sorting strategy parameters
  \State \quad // \textit{Initialize GPMA and sort structure}
  \State \quad \Call{GlobalSortParticlesByCell}{} 
  % \State \quad Initialize GPMA structures for each particle tile
  \State \quad \Call{ResetRankSortCounters}{}

  \State
  % Main PIC loop
  \State \textbf{Main PIC Loop (each timestep):}
  % \State \quad Perform field interpolation, particle push, field solve (excluding deposition)
  % \State \quad $t_{start} \leftarrow$ \Call{MPI\_Wtime}{}
  % \State \quad Increment rank-local sort step counter
  % \State \quad Initialize local \texttt{step\_stats}
  \State \quad Initialize local $step\_stats$ and $perf\_metric$

  % \State
  % Per-tile processing
  \State \quad \textbf{for particle tile $ptile$:}
  \State \quad \quad // \textit{Phase 1: Incremental Sort Preparation}
  \State \quad \quad \textbf{for cell $c_{old}$ in $ptile$:}
  \State \quad \quad \quad // \textit{SIMD by VPU}
  \State \quad \quad \quad \textbf{for particle $ip$ in cell $c_{old}$:}
  \State \quad \quad \quad \quad Prepare data for deposition 
  \State \quad \quad \quad \quad// \textit{new grid index is computed}
  % \State \quad \quad \quad \quad Compute new grid index $(i',j',k')$
  % \State \quad \quad \quad \quad Use grid index $(i',j',k')$ to compute new cell
  \State \quad \quad \quad \quad \Call{GetNewCellLocation}{$i',j',k'$}
  % \State \quad \quad \quad \quad Store data for shape‐function precomputation
  \State \quad \quad \quad \quad \textbf{if $ip$ moved to $c_{new}$}:
  \State \quad \quad \quad \quad \quad $pending\_moves$.push($ip,c_{new}$)
  \State \quad \quad \quad \textbf{end for}
  \State \quad \quad \textbf{end for}
  \State \quad \quad \Call{ApplyPendingMoves}{$ptile$, $pending\_moves$, $step\_stats$} \Comment{Rebuild GPMA if needed}
  \State \quad \quad $pending\_moves$.clear

  \State \quad \quad // \textit{Phase 2: MPU‐accelerated Deposition}
  \State \quad \quad \textbf{for cell $i_c$ in tile:}
  \State \quad \quad \quad \Call{PrepareParticleDataForMPU}{$i_c$}
  \State \quad \quad \quad $ rhocells(i_c) \leftarrow $ \Call {DepositCellCurrent}{$i_c$}
  \State \quad \quad \textbf{end for}
  \State \quad \textbf{end for} 
  % \State \quad // \textit{Phase 3: Postprocess Reduce}
  \State \quad \Call{ReduceRhocellsToGrid}{$rhocells$}

  \State
  % Reduction and global sort decision
  % \State \quad \Call{ReduceRhocellsToGlobalGrid}{all\_rhocells}
  % \State \quad $t_{end} \leftarrow$ \Call{MPI\_Wtime}{}
  % \State \quad Compute \texttt{perf\_metric} from durations and particle count
  \State \textbf{Global Sort Decision:}
  \State \quad Update local $step\_stats$ and $perf\_metric$
  \State \quad \textbf{if} \Call{ShouldPerformGlobalSort}{$step\_stats$, $perf\_metric$}:
  % \textbf{then}
  \State \quad \quad \Call{GlobalSortParticlesByCell}{}
  \State \quad \quad \Call{ResetRankSortCounters}{}
  \State \quad \textbf{end if}

\end{algorithmic}
\end{algorithm}

\subsection{Overall Algorithmic Framework}
\label{subsec:overall_framework}

The \textbf{MatrixPIC} framework is embedded within the standard PIC simulation loop, augmenting the current deposition phase. The overall workflow, particularly emphasizing the sorting and deposition stages within each MPI rank, is illustrated in Algorithm~\ref{alg:main_pic_loop_overview}.

The process begins with a one-time initialization phase per MPI rank, where particles are globally sorted by cell index using counting sort (\texttt{GlobalSortParticlesByCell}). This initial sort also initializes the GPMA structures associated with each particle tile. These structures, managed within the \texttt{ParticleTile} (or equivalent container) , include
\begin{itemize}
\item \textit{the local index array} (\texttt{m\_local\_index}), stores particle identifiers or invalid markers (e.g., \texttt{INVALID\_PARTICLE\_ID}). 
\item \textit{bin offsets} (\texttt{m\_bin\_offsets}), marks the starting index of each cell’s particles in \texttt{m\_local\_index}.
\item and \textit{lengths} (\texttt{m\_bin\_lengths}), Tracks the number of valid particles per cell.
\item \texttt{m\_num\_particles}, \texttt{m\_capacity}, \texttt{m\_num\_empty\_slots}: Manage GPMA metadata.
\item \texttt{m\_empty\_slots\_stack}: A stack tracking empty slot indices for \( O(1) \) access.
\item \texttt{m\_was\_rebuilt\_this\_step}: Flags tile rebuilding.
\end{itemize}
to delineate particles per cell within the index array, and metadata for managing GPMA capacity and empty slots. Concurrently, counters and baseline performance metrics for the global resorting strategy are reset via \texttt{ResetRankSortCounters}. User-configurable parameters governing the sorting strategy (e.g., sort interval, rebuild triggers) are also read during initialization.

Once initialized, before entering the main PIC timestep loop, the step counter of each MPI rank
for the global sort logic are incremented. After standard PIC operations like field interpolation and particle push (which updates particle positions and velocities), the current deposition phase commences. For each tile, the process is logically divided into two main phases:

\begin{enumerate}
    \item \textbf{Incremental Sort and Data Preparation (VPU-driven):}
    New particle grid indices and shape function data are computed by VPUs. Moved particles (include particles moved out or newly added of the tile) are flagged and pushed into new \texttt{pending\_moves} for subsequent updates.  After iterating through all particles in the tile, \texttt{ApplyPendingMoves} is called to move particles. These moves are applied to update the GPMA, potentially triggering local tile GPMA rebuild.
    \item \textbf{MPU-Accelerated Current Deposition (Hybrid VPU-MPU Kernel):}
    Sorted particles are processed cell-by-cell, with VPUs preparing data (such as shape factor and particle weight) and MPU-accelerated kernel performing high-density deposition and accumulate current contributions into local \texttt{rhocell} structures.
\end{enumerate}

After these two phases, the contributions stored in the local \texttt{rhocell} structures are reduced (scatter-added) to the global grid current arrays $J_x,J_y,J_z$ using VPU operations. This reduction step only requires one access per \texttt{rhocell} element for the final accumulation.

Finally, at the end of each timestep, the function \texttt{ShouldPerformGlobalSort} evaluates the collected \texttt{RankSortStats} (includes accumulated local rebuilds, global empty slot ratio across all tiles in the rank) and the current performance metric against the user-defined thresholds and the baseline performance. If any of these conditions met the sort policy, a Global sort \texttt{SortParticlesByCell} is performed for the entire rank, followed by \texttt{ResetRankSortCounters} to update the resort conditions. 

\begin{figure}[t]
  \centering
  \includegraphics[width=\linewidth]{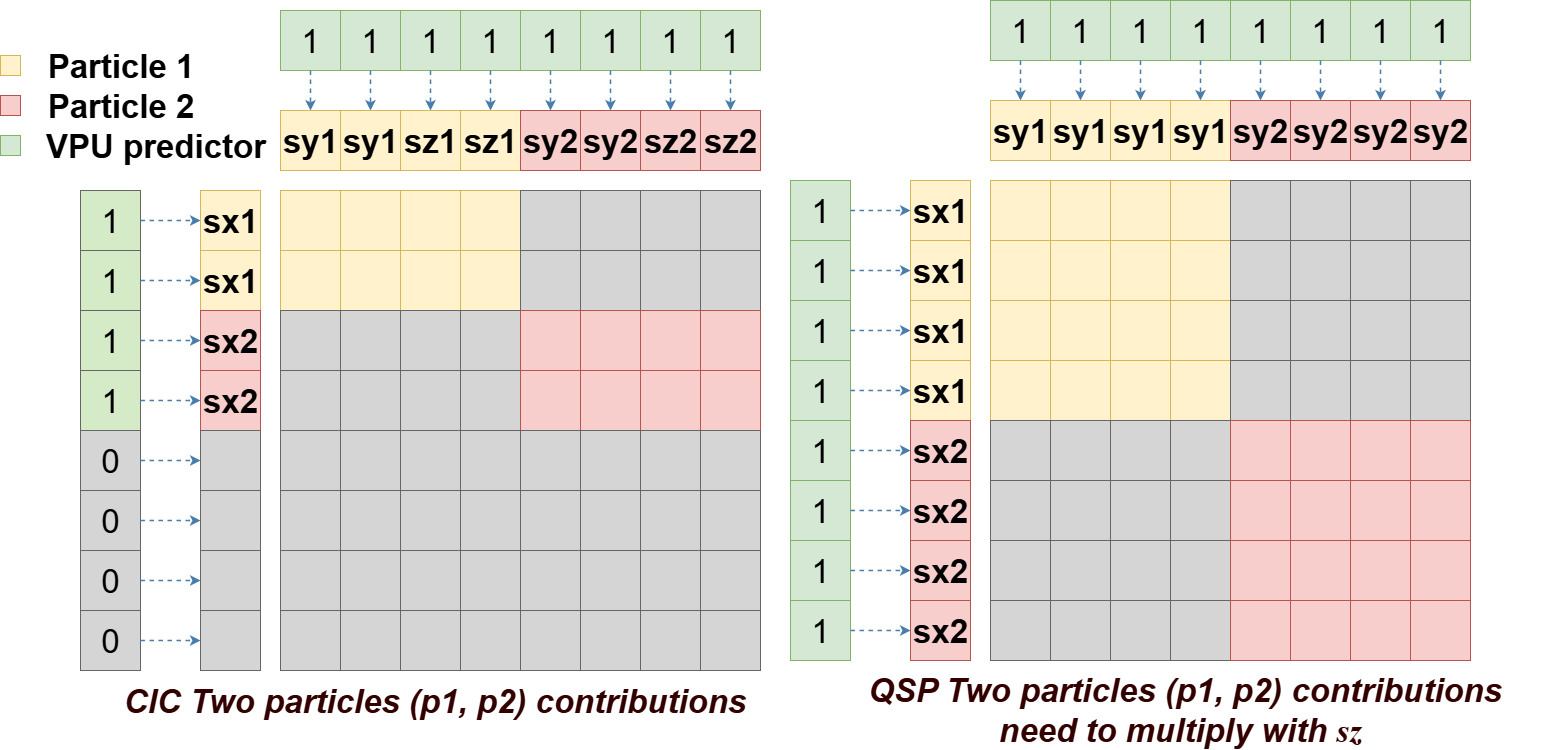} % User specified path, adjust caption as per final figure
  \caption{MPU outer product for current deposition. Left: In CIC schema, 2 particle ($p_1, p_2$) contributions (16 values total) computed within logical $4 \times 8$ outer product operation. Right: In QSP schema, 2 particle ($p_1, p_2$) contributions (32 values total) computed within logical $8 \times 8$ outer product operation.}
  \label{fig:mpu_outer_product_deposition}
\end{figure}

\subsection{Current Deposition Using MPU}
\label{subsec:mpu_deposition_kernel}

To harness the high computational density of Matrix Processing Units (MPUs), we fundamentally redesign the current deposition kernel. Our approach maps particle contributions to vector outer product operations, leveraging a \texttt{rhocell}-like data structure~\cite{Vincenti2017SIMD} for conflict-free intermediate accumulation. Then the VPU-MPU hybrid kernel intelligently combines the strengths of VPUs for data preparation and control flow with MPUs for high-throughput arithmetic.

\subsubsection{Mapping \texttt{rhocell}-based Deposition to MPU Operations}
\label{ssubsec:mapping_rhocell_mpu}

The \texttt{rhocell} concept, as introduced in Section~\ref{sec:prelim}, localizes a particle's current contributions to its immediate neighboring grid nodes, thereby mitigating atomic conflicts. For a first-order CIC scheme in 3D, each particle $p$ deposits its current to 8 surrounding grid nodes (shown in Figure~\ref{fig:rhocell_layout}). These 8 contributions (for each current component, e.g., $J_x, J_y, J_z$) are stored in the \texttt{rhocell} (Equation~\ref{eq:rhocells}).

For a given cell $c$, the update to the $k$-th node entry in its \texttt{rhocell} by particle $p$ with effective current contribution $w_{p,x} = q_p v_{p,x} W_p$ (where $W_p$ is the particle's macro-particle weight) and combined 3D shape function $S_k(\mathbf{x}_p)$ is:
\begin{equation}
\texttt{rhocell}[c][k] \leftarrow \texttt{rhocell}[c][k] + w_{p,x} S_k(\mathbf{x}_p).
\label{eq:rhocell_single_particle_update}
\end{equation}
The 3D shape function $S_k(\mathbf{x}_p)$ is a product of 1D shape factors, e.g., $S_{ijk}(\mathbf{x}_p) = s_{x,i}(dx_p) s_{y,j}(dy_p) s_{z,k}(dz_p)$, where $(dx_p, dy_p, dz_p)$ are the particle's normalized intra-cell coordinates.

% Placeholder for a pseudocode for the hybrid deposition kernel
\begin{algorithm}[t]
\caption{Hybrid VPU-MPU Current Deposition Kernel for a Cell}
\label{alg:hybrid_deposition_kernel}
\begin{algorithmic}[1]
\Require Particles in the current cell $ic$ (accessed via GPMA indices), global grid $J$
\Ensure Current in global grid $J$ is updated.
% \Ensure Updated local \texttt{rhocell} structures for particles in this cell.
\State \textit{// Stage 1: VPU Preprocessing (per VPU vector of particles)}
\State Load particle positions $\mathbf{x}_p$, velocities $\mathbf{v}_p$, charges $q_p$, weights $W_p$ into VPU registers.
\State \textbf{for each particle $p$ in the VPU vector}:
\State \quad Compute integer grid indices $(ix_p, iy_p, iz_p)$ and normalized intra-cell coordinates $(dx_p, dy_p, dz_p)$.
\State \quad Update GPMA sort index by \Call{GetCellID}{$ix_p, iy_p, iz_p$}.
\State \quad Compute 1D shape function pairs: $(s_{x,0}, s_{x,1})$, $ (s_{y,0}, s_{y,1})$, $ (s_{z,0}, s_{z,1})$.
\State \quad Compute weights: $wqx, wqy, wqz$.
\State \quad Store precomputed values (e.g., $wqx, wqy, wqz$, $s_x, s_y, s_z$, $(ix_p, iy_p, iz_p)$) into temporary arrays.
\State \textbf{end for}
\State
\State \textit{// Stage 2: MPU Deposition (iterate in batches)}
\State \textbf{for each batch of SORTED $\ell$ particles from the preprocessed data do}
\State \quad \textit{// VPU: Prepare input vectors $\mathbf{A}$ and $\mathbf{B}$ for MPU}
\State \quad Assemble vector $\mathbf{A}$ and $\mathbf{B}$ \Comment{Section \ref{ssubsec:mapping_rhocell_mpu}}
\State \quad Load $\mathbf{A}$ and $\mathbf{B}$ into MPU input vector registers.
\State \quad \textit{// MPU: Execute Outer Product}
\State \quad $\mathbf{C}_{\text{tile}} \leftarrow \mathbf{A} \otimes \mathbf{B}$ \Comment{remain in register}
\State \textbf{end for}
\State \textit{// VPU: Extract results from MPU tile and store to \texttt{rhocell}}
\State \textbf{for each row in $\mathbf{C}_{\text{tile}}$ do}
% \State \quad Read out from $\mathbf{C}_{\text{tile}}$ (VPU row reads from MPU tile).
\State \quad Scatter contributions into the proper \textit{rhocells($ic$)}.
\State \textbf{end for}
\State
\State \textit{// Stage 3: VPU Postprocessing (see Algorithm~\ref{alg:main_pic_loop_overview})}
\State \textbf{for each row in \texttt{rhocell} do}
\State \quad Reduce contribution to the global grid $\mathbf{J}$ using VPU gather/scatter-add.
\State \textbf{end for}
\end{algorithmic}
\end{algorithm}

A straightforward way to compute the 8 nodal contributions for a single particle $p$ using an MPU outer product is by constructing two vectors:
\begin{itemize}
    \item $\mathbf{a} = [s_{x,0}, s_{x,1}]^T \in \mathbb{R}^2$, representing the $x$-direction shape factors for particle $p$.
    \item $\mathbf{b} = [s_{y,0}s_{z,0}, s_{y,1}s_{z,0}, s_{y,0}s_{z,1}, s_{y,1}s_{z,1}]^T \in \mathbb{R}^4$, combining the $y$- and $z$-direction shape factor products for particle $p$.
\end{itemize}
The MPU then computes the outer product $\mathbf{C}_{2 \times 4} = \mathbf{a} \otimes \mathbf{b}$. The resulting $2 \times 4$ matrix contains the 8 necessary 3D shape function products $S_k(\mathbf{x}_p)$. This matrix is subsequently scaled by the particle's effective current $w_{p,x}$ and its elements are flattened to update the 8 entries of the particle's $J_x$ \texttt{rhocell}:
\begin{equation}
\Delta \texttt{rhocell}_{J_x,p} \leftarrow \text{flatten}(w_{p,x} \cdot (\mathbf{a} \otimes \mathbf{b})).
\label{eq:single_particle_outer_product}
\end{equation}
While this formulation correctly maps the deposition to an outer product, a $2 \times 4$ operation underutilizes typical MPUs designed for larger tiles (e.g., $8 \times 8$ for double-precision elements, corresponding to vector operand lengths of $\ell=8$).

To maximize MPU register utilization and enhance computational density, we extend this operation to process two particles, $p_1$ and $p_2$, simultaneously using a single, larger outer product. We construct the input vectors as follows:
\begin{itemize}
    \item Let $\mathbf{a}_1 = [s_{x,0}^{(p_1)} w_{p_1,x}, s_{x,1}^{(p_1)} w_{p_1,x}]^T \in \mathbb{R}^2$ and 
    \\$\mathbf{a}_2 = [s_{x,0}^{(p_2)} w_{p_2,x}, s_{x,1}^{(p_2)} w_{p_2,x}]^T \in \mathbb{R}^2$, \\where $w_{p_i,x} = q_{p_i} v_{p_i,x} W_{p_i}$ is the scaled charge-velocity product for particle $p_i$. We then form the combined vector $\mathbf{A} = \text{concat}(\mathbf{a}_1, \mathbf{a}_2) \in \mathbb{R}^4$.
    \item Similarly, let 
    \\$\mathbf{b}_1 = [s_{y,0}^{(p_1)}s_{z,0}^{(p_1)}, s_{y,1}^{(p_1)}s_{z,0}^{(p_1)}, s_{y,0}^{(p_1)}s_{z,1}^{(p_1)}, s_{y,1}^{(p_1)}s_{z,1}^{(p_1)}]^T \in \mathbb{R}^4$ \\for particle $p_1$, and $\mathbf{b}_2 \in \mathbb{R}^4$ be the corresponding vector for particle $p_2$. We form the combined vector \\$\mathbf{B} = \text{concat}(\mathbf{b}_1, \mathbf{b}_2) \in \mathbb{R}^8$.
\end{itemize}
The MPU computes the outer product $\mathbf{C}_{MPU} = \mathbf{A} \otimes \mathbf{B}$, resulting in a $4 \times 8$ matrix tile. This single MPU operation computes the 8 nodal contributions for particle $p_1$ (found in the submatrix formed by the first two rows of $\mathbf{C}_{MPU}$ interacting with the first four columns, i.e., $\mathbf{a}_1 \otimes \mathbf{b}_1$) and the 8 nodal contributions for particle $p_2$ (found in the submatrix formed by the last two rows of $\mathbf{C}_{MPU}$ interacting with the last four columns, i.e., $\mathbf{a}_2 \otimes \mathbf{b}_2$). The resulting cross-terms from $\mathbf{a}_1 \otimes \mathbf{b}_2$ and $\mathbf{a}_2 \otimes \mathbf{b}_1$ are not directly used and can be zeroed out during operand construction. This method effectively computes $8+8=16$ contribution terms within one $4 \times 8$ logical outer product, efficiently utilizing a portion of a larger MPU tile and doubling the computational density for these two particles compared to processing them with separate $2 \times 4$ outer products or sequential VPU operations. This process is conceptually illustrated in Figure~\ref{fig:mpu_outer_product_deposition}. 

Furthermore, this outer product formulation can be extended to higher-order deposition schemes like second-order Triangular Shaped Cloud (TSC) or third-order Quadratic Spline (QSP). For instance, with a third-order QSP scheme, each 1D shape function involves 4 terms (e.g., $s_{x,0}, s_{x,1}, s_{x,2}, s_{x,3}$), leading to $4^3 = 64$ nodal contributions per particle in 3D. A similar strategy of combining shape factors from different dimensions into MPU vector operands can be employed. For example, for two particles ($p_1, p_2$) and QSP, one could form an MPU input vector $\mathbf{A} \in \mathbb{R}^8$ by concatenating the scaled $s_x$ terms for $p_1$ and $p_2$ (i.e., 
\[
[w_{p1,x}s_{x,0}^{(p1)}, \dots, w_{p1,x}s_{x,3}^{(p1)}, w_{p2,x}s_{x,0}^{(p2)}, \dots, w_{p2,x}s_{x,3}^{(p2)}]^T
\]
). Another input vector $\mathbf{B} \in \mathbb{R}^8$ could be formed by concatenating the $s_y$ terms for $p_1$ and $p_2$ (i.e., 
\[
[s_{y,0}^{(p1)}, \dots, s_{y,3}^{(p1)}, s_{y,0}^{(p2)}, \dots, s_{y,3}^{(p2)}]^T
\]
). The $8 \times 8$
outer product $\mathbf{A} \otimes \mathbf{B}$ would yield a $64$-element matrix where each element is a product of a scaled $s_x$ term and an $s_y$ term (as shown in Figure~\ref{fig:mpu_outer_product_deposition}). This intermediate $64$-element result (logically two $4 \times 4$ blocks for each particle) then needs to be multiplied element-wise by the corresponding $s_z$ terms (4 for each particle) and accumulated. This final multiplication by $s_z$ terms and accumulation into the 64-node \texttt{rhocell} could potentially be handled by VPUs. This approach could effectively utilize up to $8 \times 8 = 64$ MPU tile size ($4 \times 8 = 32$ elements are valid) for the $s_x \cdot s_y$ part of the calculation for two particles, indicating the potential for even higher compute density with higher-order schemes. While this paper focuses on demonstrating the performance effects with first-order CIC deposition, the proposed MPU mapping strategy provides a viable path for higher-order schemes.

\subsubsection{Hybrid VPU-MPU Kernel Design}
\label{ssubsec:hybrid_kernel}
The current deposition kernel is meticulously hand-crafted using low-level intrinsics to orchestrate VPU and MPU operations, minimizing overhead and maximizing data reuse. The workflow for depositing current for particles within a given cell, after they have been sorted and their raw data (positions, weights and etc.) is available, proceeds in stages as outlined in Algorithm~\ref{alg:hybrid_deposition_kernel}:

\paragraph{Stage 1: VPU Preprocessing and Data Staging:}
As described in Section~\ref{subsec:overall_framework} (Phase 1 of per-tile processing), for particles within a cell (accessed in their sorted order), VPUs are first responsible for all preparatory computations. This includes loading raw particle data (positions, velocities, charge $q_p$, weight $W_p$ and etc.), calculating their logical tile grid indices $(ix_p, iy_p, iz_p)$ and normalized intra-cell coordinates $(dx_p, dy_p, dz_p)$. From these, the fundamental 1D shape function components (e.g., $s_{x,0} = 1-dx_p, s_{x,1} = dx_p$, and similarly for $y, z$) are computed. These calculations are performed using VPU vector instructions (e.g., processing 8 double-precision particles for a 512-bit VPU at one time). The resulting values, such as the particle current term $\text{Wqx}_{p} = q_p v_{p,x} W_p$ and the shape factor $s_x, s_y, s_z$ are stored in temporary 1D arrays. This phase also handles any conditional logic, such as applying boundary conditions or checking particle validity, leveraging the VPU's general-purpose instruction set.

\paragraph{Stage 2: MPU-Dominated Deposition to \texttt{rhocell}}
Once the necessary per-particle terms and shape factors are prepared by the VPU, they are formatted into vectors suitable for MPU outer product operations as described in Section~\ref{ssubsec:mapping_rhocell_mpu}. For two particles \(p_1\) and \(p_2\), the VPU concatenates their weights \(w_{p_i,x} \cdot s_{x,k}^{(p_i)}\) (for \(k=0,1\)) into a vector \(\mathbf{A} \in \mathbb{R}^4\), and computes and concatenates \(s_{y,k}^{(p_i)} \cdot s_{z,m}^{(p_i)}\) into \(\mathbf{B} \in \mathbb{R}^8\). These vectors are fed into the MPU, which computes the outer product \(\mathbf{C_{tile}} = \mathbf{A} \otimes \mathbf{B} \in \mathbb{R}^{4 \times 8}\). Due to the cell-sorted order of particles (detailed in Section~\ref{subsec:sorting_strategy}), contributions for all particles within the same cell are processed consecutively. This allows the MPU tile register ($\mathbf{C}_{\text{tile}}$) to remain resident in the MPU's fast memory across multiple outer product operations for different pairs of particles within that same cell. This significantly reduces data movement between the MPU and VPU, as the tile is only written out to the \texttt{rhocell} structure for that cell once all its particles have been processed. To further enhance instruction-level parallelism and saturate the MPU's pipeline, we unroll the MPU outer product computations. For instance, two MPU pipelines can be employed concurrently to process four particles (two pairs) simultaneously.
After all particles within a cell have been processed, the VPU reads out the data row by row from the MPU tile. Only the contributions corresponding to valid particles are selected and then reduced by the VPU and written to the corresponding cell's final \texttt{rhocell} array in memory.

\paragraph{Stage 3: VPU-based \texttt{rhocell} Reduction (Post-processing):}
After processing all particles across cells in a rank, a VPU-driven pass reduces the \texttt{rhocell} contributions to the global current density arrays (\(\mathbf{J}_x, \mathbf{J}_y, \mathbf{J}_z\)). As noted by Vincenti et al.~\cite{Vincenti2017SIMD}, each \texttt{rhocell}’s contiguous 8-element vector requires scattering to the corresponding grid nodes. We implement efficient VPU-based indexed scatter operations to write \texttt{rhocell} data to \(\mathbf{J}_x\), \(\mathbf{J}_y\), and \(\mathbf{J}_z\), with vector unrolling and optimized instruction scheduling to enhance performance.

This hybrid VPU-MPU intrinsic-based kernel is the cornerstone of \textbf{MatrixPIC}'s strategy to leverage MPU computational power. It carefully balances the workload: VPUs handle complex control flow, data preparation, and final data marshalling, while MPUs are dedicated to the high-density arithmetic of accumulating current contributions via outer products. The effectiveness of this stage heavily relies on the particle sorting (detailed in Section~\ref{subsec:sorting_strategy}) to ensure MPU registers reuse as much as possible and minimizing costly spills to VPU registers.

\subsection{Efficient Incremental Particle Sorting using GPMA}
\label{subsec:sorting_strategy} % Changed from sorting_strategy to match content better

A key enabler for maximizing the MPU's computational efficiency and register reuse in our \textbf{MatrixPIC} framework is the maintenance of particle data in a cell-sorted order. However, performing a full particle sort at every timestep is computationally prohibitive. To address this, we introduce an efficient incremental particle sorting strategy that leverages the GPMA for low-overhead index management. This approach is motivated by the CFL condition in PIC simulations, which implies that most particles do not cross cell boundaries in a single timestep~\cite{bib_CFL_condition_PIC_ref}.

\begin{figure}[ht]
  \centering
  \includegraphics[width=1\linewidth]{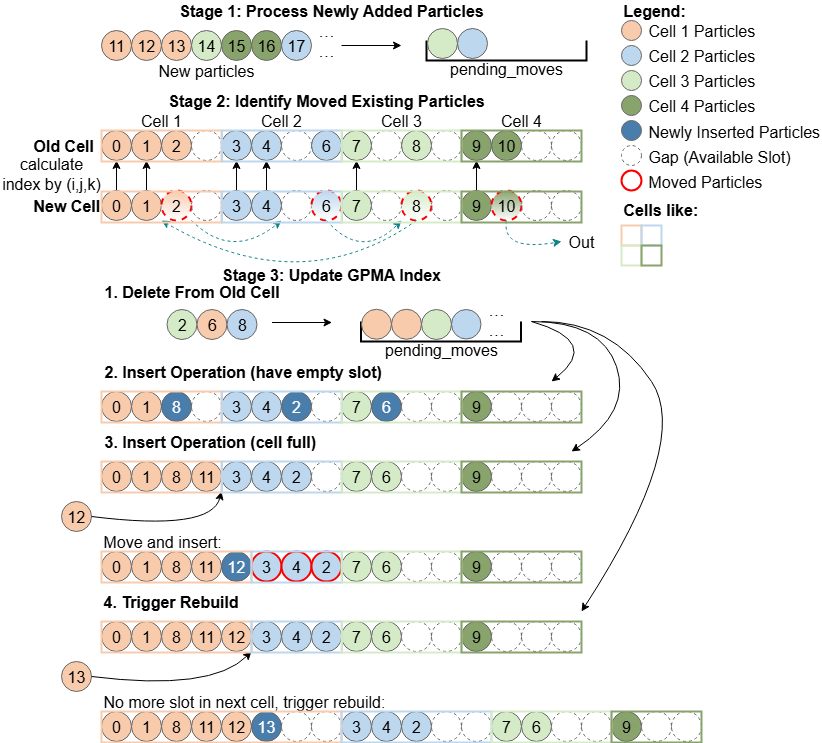}
  \caption{Incremental Particle Sorting within a Particle Tile using GPMA.}
  \label{fig:increment_sort}
\end{figure}

Our incremental sorting mechanism is tightly integrated into the current deposition kernel within the VPU-driven preprocessing phase (Stage 1 of Algorithm~\ref{alg:hybrid_deposition_kernel}). The core idea is to only update particles that have actually moved to a new grid cell, while leaving the indices of stationary particles untouched. The overall process is depicted in Figure~\ref{fig:increment_sort}.

\subsubsection{Incremental Particle Sorting Algorithm}
\label{ssubsec:incremental_sorting_algo}
After the global initialization phase where all particles are sorted and their corresponding GPMA structures (e.g., \texttt{m\_local\_index}, \texttt{m\_bin\_offsets}, \\\texttt{m\_bin\_lengths}, \texttt{m\_num\_empty\_slots}, and an \texttt{m\_empty\_slots\_stack} within each particle tile's \texttt{ptile} data structure) are populated, the per-timestep incremental sort proceeds as follows within the VPU preprocessing stage of the current deposition kernel:

\paragraph{Stage 1: Processing Newly Added Particles:} Newly added particles are processed first. For each new particle, its grid cell is determined, and it is added to a temporary list, \texttt{m\_pending\_moves}, flagged for insertion into the appropriate GPMA bin.

\paragraph{Stage 2: Identifying Moved Existing Particles:} The algorithm then iterates through each cell $c_{old}$ within the particle tile, accessing particles via the sorted indices in \texttt{m\_local\_index}. This iteration is performed using VPU vectorization. For each valid particle $p$:
\begin{enumerate}
    \item Its updated position is used to determine its new grid cell, $c_{new}$. This reuses the position data already computed by pre-process phase.
    \item If $c_{new}$ is different from its current cell $c_{old}$, the particle is marked as moved. Its ID $p$ and new cell $c_{new}$ are added to the \texttt{m\_pending\_moves} list.
    \item Concurrently, the particle's original slot in \texttt{m\_local\_index} corresponding to $c_{old}$ is marked as \texttt{INVALID\_PARTICLE\_ID}, effectively deleting it from its old bin. The metadata \\\texttt{m\_bin\_lengths[$c_{old}$]} and the tile's \texttt{m\_num\_particles} are decremented, while \texttt{m\_num\_empty\_slots} is incremented and the now-empty slot index is pushed onto \\\texttt{m\_empty\_slots\_stack}.
\end{enumerate}
This entire process is efficiently performed using VPU vector instructions.

After scanning all particles in the tile, the \texttt{m\_pending\_moves} list contains all particles that need to be inserted into new bins (either newly added particles or existing particles that moved). The next stage involves efficiently updating the GPMA index structure using this list.

\subsubsection{Efficient Sort Index Update Using GPMA}
\label{ssubsec:gpma_update}
The GPMA is employed to manage the \texttt{m\_local\_index} array within each particle tile, enabling highly efficient updates to the sorted particle indices~\cite{Bender2007AdaptivePMA, Durand2012PMA_MovingParticles}. Traditional dense arrays make insertions and deletions costly due to large data shifts, while linked-lists suffer from poor traversal performance for SIMD operations. GPMA strikes a balance by introducing "gaps" (empty slots marked by \texttt{INVALID\_PARTICLE\_ID} and tracked by \texttt{m\_empty\_slots\_stack}) within contiguous blocks of indices representing particle bins (Figure~\ref{fig:gpma}).

The update phase, \texttt{ApplyPendingMovesAndRuildGPMA}, processes the \texttt{m\_pending\_moves} list:
\begin{itemize}
    \item \textbf{Deletion:} As particles are identified as having moved \emph{from} a cell, their original entries in \texttt{m\_local\_index} are marked as empty (\texttt{INVALID\_PARTICLE\_ID}). The indices of these now-empty slots are pushed onto the \texttt{m\_empty\_slots\_stack} for the tile. This is an $O(1)$ operation. The corresponding \texttt{m\_bin\_lengths} and \texttt{m\_num\_particles} are updated.

    \item \textbf{Insertion:} For each particle $p$ that needs to be inserted into a target cell $c_{new}$:
    \begin{enumerate}
        \item The primary mechanism for insertion is to utilize an available empty slot within the target cell's bin (or a globally available empty slot within the tile's \\\texttt{m\_local\_index} if bins are allowed to dynamically grow into general free space). An empty slot index is popped from the tile's \texttt{m\_empty\_slots\_stack}. The particle ID $p_{id}$ is placed at this index in \\\texttt{m\_local\_index}. This effectively adds the particle to the logical list for cell $c_{new}$. This operation is $O(1)$, assuming the stack is not empty. The \texttt{m\_bin\_lengths[$c_{new}$]} is incremented, and \\\texttt{m\_num\_empty\_slots} is decremented.
        
        \item If \texttt{m\_empty\_slots\_stack} for this cell is empty, it indicates that cell $c_{new}$ is logically full. It will first try to borrow empty slot from next bin $c_{next}$. This involves finding an empty slot in $c_{next}$. Then, a block of particle indices from the end of $c_{new}$'s bin up to (but not including) the borrowed slot in $c_{next}$'s bin must be shifted by one position to create a contiguous space for $p$ at the logical end of $c_{new}$'s bin. This shifting operation has a cost proportional to the number of elements moved, potentially under $O(N_{cell\_capacity})$ in a localized worst case.

        If borrowing is not feasible (e.g., adjacent bins are also full, or other conditions triggered), it signifies that the GPMA structure for the \emph{entire tile} requires reorganization. A rebuild of the GPMA structure is triggered if: 
        (1) insertion fails (non-empty overflow list), 
        (2) global empty slots drop below a threshold, or 
        (3) overflow particles exceed a limit. 
        It is mandatory when overflow particles exist with critically low empty slots, or if overflow is excessive; otherwise, it is optional for low empty slots.
    \end{enumerate}
    \item \textbf{GPMA Local Rebuild:} This rebuild (complexity $O(N_{p,tile})$) re-allocates \texttt{m\_local\_index} for the tile, potentially increasing its capacity (\texttt{m\_capacity}), re-distributes all valid particles contiguously within their respective cell bins, and re-establishes a new set of uniformly distributed gaps and a fresh \texttt{m\_empty\_slots\_stack}. The flag \\\texttt{m\_was\_rebuilt\_this\_step} is set for this tile, and the count of local rebuilds is updated for the global sort strategy.
\end{itemize}

In summary, this two-tiered approach—VPU-driven identification of moved particles and GPMA-based efficient index updates—provides a low-overhead incremental sorting mechanism. It maintains the cell-sorted particle order required for MPU efficiency. Since GPMA does not alter the order of particles in memory, we therefore need a global reorganization strategy, which will be introduced in Section~\ref{subsec:global_resorting_policy}.

\subsection{Global Re-sorting Policy}
\label{subsec:global_resorting_policy}

The global re-sorting policy mitigates long-term inefficiencies in GPMA structures caused by particle migration and density variations across MPI ranks. It adaptively adjusts GPMA to reduce tile-level rebuilds and enhances memory coherence for cell-adjacent particles, improving VPU access continuity. The policy is triggered post-PIC cycle via \texttt{ShouldPerformRankLocalSort}, based on five prioritized strategies:

\begin{enumerate}
    \item \textbf{Minimum Interval}: Skips sorting if steps since the last sort are below \texttt{m\_min\_sort\_interval}.
    \item \textbf{Fixed Interval}: Triggers every \texttt{m\_sort\_interval} steps.
    \item \textbf{Local Rebuilds}: Triggers if cumulative GPMA rebuilds exceed \texttt{m\_sort\_trigger\_rebuild\_count}.
    \item \textbf{Empty Slot Ratio}: Triggers if empty slots drop below \\\texttt{m\_sort\_trigger\_empty\_ratio} or over \\\texttt{m\_sort\_trigger\_full\_ratio}.
    \item \textbf{Performance Degradation (optional)}: Triggers if performance falls below \texttt{m\_sort\_trigger\_perf\_degrad} of baseline.
\end{enumerate}

Upon triggering, \texttt{GlobalSortParticlesByCell} employs counting sort to reorder particles and rebuild GPMA, followed by \texttt{ResetRankSortCounters} to reset metrics. These strategies are user-configurable for adaptability.
The parameters governing these sorting policies are user-configurable, allowing tuning for different simulation scenarios and hardware characteristics.

\section{Experimental Setup}
\label{sec:Experimental_Setup}

\subsection{Platform Details}

\begin{figure}[h]
  \centering
  % Please replace 'path/to/your/figure.png' with the actual path to your architecture diagram.
  \includegraphics[width=\linewidth]{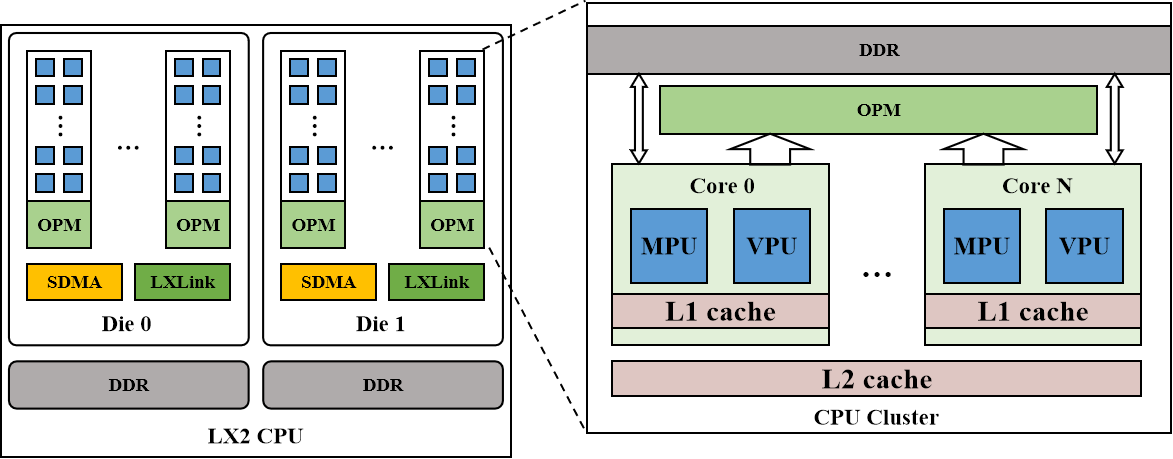} 
  \caption{Architectural diagram of the LX2 processor, illustrating the dual-Die design, core distribution, and key components like the VPU, MPU, and memory interfaces.}
  \label{fig:lx2_arch}
\end{figure}

Our primary experiments were conducted on the LS pilot system, a next-generation HPC cluster. Each compute node incorporates two LX2 high-performance CPUs. As illustrated in the architectural diagram in Figure~\ref{fig:lx2_arch}, each processor package contains over 256 cores distributed across two computing Dies.

These cores support both vector (VPU) and matrix (MPU) computation engines. The VPUs execute double-precision (FP64) SIMD instructions, while the MPUs are designed for 8$\times$8 matrix operations. Critically, the MPU's MOPA instruction offers approximately 4$\times$ the theoretical FP64 performance of the VPU's Multiply-Accumulate (MLA) instruction, presenting a significant opportunity for acceleration. Both compute units operate at frequencies at or above 1.3 GHz.

The LX2 architecture implements a system-on-chip design where each Die features 128GB of off-die DDR memory organized across four NUMA domains. To enhance data movement efficiency between DDR and the high-bandwidth on-package memory (OPM), a System Direct Memory Access (SDMA) interface is implemented in each CPU Die. The Dies are interconnected through an LXLink network, with each network card providing a bidirectional bandwidth of 48 GB/s.

The system employs a customized Linux-based OS with an optimized software stack comprising:
\begin{itemize}
    \item LLVM-based Clang/Flang compilers with architecture-specific auto-vectorization.
    \item A tuned OpenMPI implementation with optimized collective operations.
    \item High-performance math libraries (BLAS, LAPACK, FFT, etc.) with architecture-specific optimization.
\end{itemize}

For cross-platform comparison, we also conducted experiments on a platform equipped with two 28-core CPUs and 8 NVIDIA A800 GPUs, which features 80 GB of HBM2e memory. The software environment used for these tests included NVIDIA driver version 535.104.12.
% For cross-platform comparison, we also conducted experiments on the Tianhe Xingyi cluster. Each compute node in this system is equipped with two 28-core CPUs and eight NVIDIA A800 GPUs. The nodes feature 1024 GB of shared system memory. Each A800 GPU is equipped with 80 GB of HBM2e memory. The software environment used for these tests included NVIDIA driver version 535.104.12.

\subsection{Evaluation Methodology}
\label{subsec:evaluation_methodology}

Our experimental evaluation utilizes \texttt{WarpX}~\cite{WarpX2022} version 24.07, a highly optimized PIC code from Lawrence Berkeley National Laboratory. All simulations are compiled with \texttt{-O3} and \texttt{-flto} optimizations and use the CKC Maxwell solver and the Boris particle pusher. Each configuration is run for 100 steps, with each experiment repeated three times for statistical reliability. The particle density is scaned across a range of particle-per-cell (PPC) values from 1 to 128 (PPC = 1, 4, 8, 16, 32, 64, 128). We employ two primary workloads: a \textit{uniform plasma} simulation (using first-order CIC and third-order QSP schemes) for controlled performance analysis, and a \textit{Laser-Wakefield Acceleration (LWFA)} simulation (using the CIC scheme) to assess performance in a realistic application~\cite{Tajima1982Laser, Esarey1996Overview}.

The core parameters defining each workload and the adaptive sorting policy for \textbf{MatrixPIC} are summarized in Appendix~\ref{appen:parameter}, Table~\ref{tab:workload_parameters}.

% To systematically analyze our contributions, we perform two sets of ablation studies on the uniform plasma workload. The first set dissects the components of our framework:
\subsubsection{Evaluation Setup}
To systematically evaluate our framework, we conduct both an \textbf{ablation study} to dissect its internal components and a \textbf{comparative study} against state-of-the-art VPU-based methods. The first set dissects the components of our framework:
\begin{description}
    \item [\texttt{Baseline}] The unmodified \texttt{WarpX} kernel, serving as the performance reference.
    \item [\texttt{Matrix-only}] An MPU-only kernel to isolate its raw computational performance.
    \item [\texttt{Hybrid-noSort}] The hybrid MPU-VPU kernel without sorting to assess its impact.
    \item [\texttt{Hybrid-GlobalSort}] The hybrid kernel with a non-incremental global sort each timestep for comparison.
    \item [\texttt{FullOpt (MatrixPIC)}] The complete framework with all proposed optimizations integrated.
\end{description}

The second set provides a rigorous comparison against progressively stronger VPU-based baselines:
\begin{description}
    \item[\texttt{Baseline+IncrSort}] The baseline kernel enhanced with our incremental sorting algorithm to quantify its standalone benefit.
    \item[\texttt{Rhocell}] A reproduction of the compiler-vectorized \texttt{rhocell} implementation, a strong community-standard baseline.
    \item[\texttt{Rhocell+IncrSort}] The \texttt{rhocell} baseline enhanced with our incremental sorting algorithm.
    \item[\texttt{Rhocell+IncrSort (VPU)}] A manually vectorized version of the \texttt{Rhocell+IncrSort}, creating the strongest possible VPU-based competitor.
\end{description}
Note that these VPU-focused studies were performed at 128 particles-per-cell (ppc).

\subsubsection{Evaluation Metrics}
To evaluate these configurations, we measure several key metrics after a warm-up phase. Primary metrics include \textbf{Wall Time} (average execution time per step) and the complete \textbf{Deposition Kernel Time}, which encompasses all associated data preparation, sorting, and reduction steps. From these timings, we derive kernel throughput as \textbf{Particles per Second}($N_{\text{particles}} / T_{\text{deposition}}$), and relative performance as \textbf{Speedup} ($T_{\text{baseline}} / T_{\text{optimized}}$). 

% FOM
% For a more principled and application-centric cross-platform comparison, we adopt the \textbf{Figure of Merit (FOM)} used by the \texttt{WarpX} team for the U.S. Department of Energy's Exascale Computing Project (ECP)~\cite{WarpX2022}. This FOM is designed to measure the time-to-solution for a representative, scalable scientific problem. The general form of the WarpX FOM is given by:
% \begin{equation}
%     \text{FOM} = \frac{\alpha N_c + \beta N_p}{\text{avg. time per step} \times \text{percent of system used}}
%     \label{eq:fom}
% \end{equation}
% where $N_c$ is the number of cells, $N_p$ is the number of particles, and the weights are set to $\alpha =0.1$ and $\beta=0.9$. For our single-node saturation tests, we adapt this metric by setting the \textit{percent of system used} to 1 and using the Deposition Kernel Time (inclusive of all implementation-specific overheads like sorting and initialization) as the denominator. To ensure saturation on both platforms, we first fix a high particle density of $PPC=512\ (8\times8\times8)$ then scale the grid size ($N_c$) on each machine until its respective memory is exhausted (main memory for the CPU, HBM for the GPU). This adapted FOM provides a fair, application-centric comparison, effectively measuring the throughput of the deposition kernel when processing a hardware-saturating workload.

%% 0831 Percentage of Theoretical Peak Performance version
For cross-platform comparison—where raw times can be misleading—we also report \textbf{Percentage of Theoretical Peak Performance}. To calculate this, we define the \textit{effective computational work} using the canonical scalar deposition algorithm and exclude optimization-specific overheads (e.g., sorting and data initialization). For the third-order QSP scheme, this amounts to 419 floating-point operations per particle. The kernel execution time, however, is measured inclusively of all implementation-specific costs, including data structure initialization overheads. On the LX2 CPU, we saturate a single compute node with a high-density workload of $PPC=512\ (8\times8\times8)$. On the GPU, we use the same PPC and scale the grid size to the limits of its memory capacity. The resulting effective FLOP/s is then used to calculate the percentage of the hardware's theoretical peak. This methodology provides a fair measure of efficiency by crediting each implementation only for the essential scientific work while penalizing it for all associated overheads.

% This methodology reports how efficiently each implementation exploits the available architectural potential, independent of platform-specific overheads.
% Finally, for a more principled cross-platform comparison, where direct time comparisons can be misleading, we assess the \textbf{Percentage of Theoretical Peak Performance}. This metric measures the ratio of sustained kernel performance to the hardware's theoretical peak, indicating how efficiently an implementation exploits the available architectural potential.

\section{Evaluation Results}
\label{sec:Results}
\subsection{Overall Performance}

\begin{figure}[t]
\centering
\includegraphics[width=\linewidth]{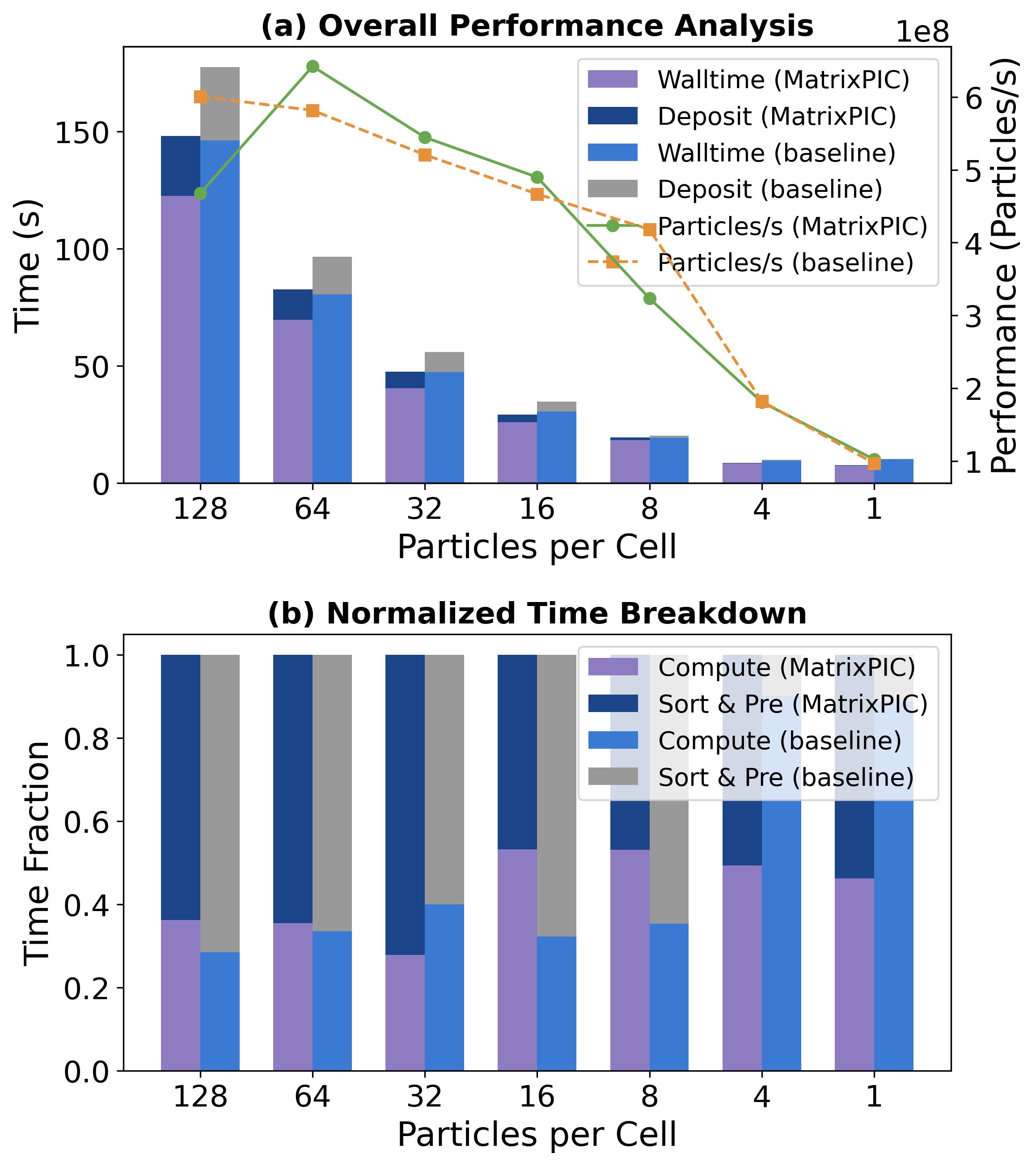}
\caption{Overall performance for the Uniform Plasma workload across varying PPC densities.
(a) Comparison of stacked execution time (total wall time and deposition kernel components) and kernel throughput (Particles/s).
(b) Normalized time breakdown illustrating the fraction of time spent in the compute kernel versus overhead from sorting and precomputation.}
\label{fig:overall}
\end{figure}

Figure~\ref{fig:overall} illustrates the overall performance of \textbf{MatrixPIC} against the baseline \texttt{WarpX} on the uniform plasma workload. At high density (PPC=128), \textbf{MatrixPIC} achieves a \textbf{16.2\%} speedup in total wall time and processes \textbf{22.13\%} more particles per second. The speedup for the deposition kernel itself is even more pronounced, reaching up to \textbf{36.4\%} at PPC=32. This result is particularly noteworthy given that the baseline is already highly optimized. However, the framework's overheads are not fully amortized at very low particle densities, where performance can be up to 17.2\% lower than the baseline (at PPC=1).

A deeper analysis reveals further potential. Since \textbf{MatrixPIC} processes 16 valid grid points per MPU instruction versus 8 for the auto-vectorized baseline, a nearly 2$\times$ kernel speedup might be anticipated. While the observed ~1.5$\times$ speedup confirms the effectiveness of our strategies, it also indicates that performance is influenced by data movement between the VPU and MPU, intrinsic latencies, and other VPU-bound operations in the hybrid kernel. Further tuning of data tiling and sorting heuristics could help close this gap and more fully realize the MPU's theoretical advantage.

\begin{figure}[t]
  \centering
  \includegraphics[width=\linewidth]{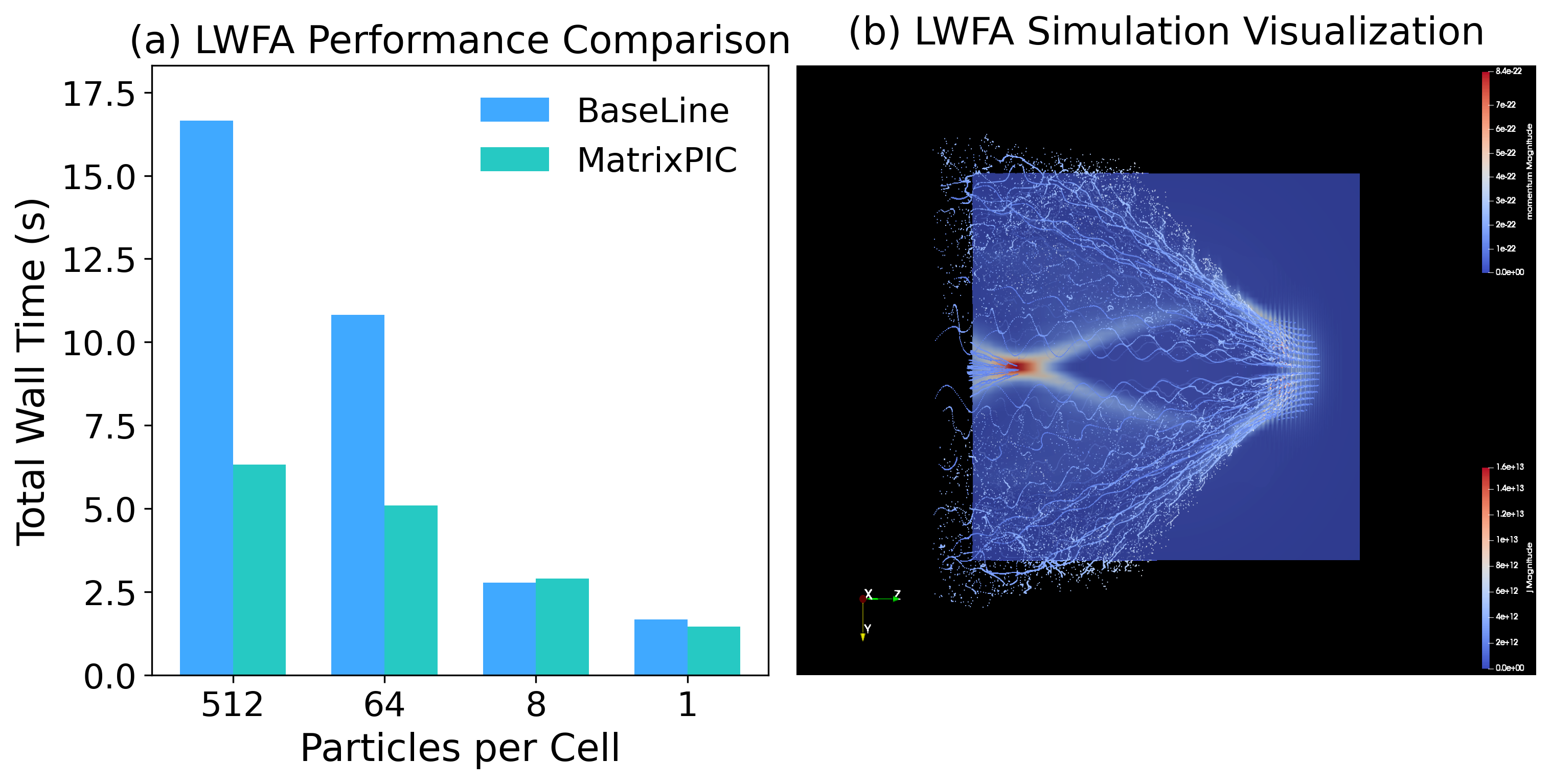}
  \caption{Comparison between \texttt{MatrixPIC} and \texttt{WarpX} in LWFA: (a) Total wall time across variant ppc (b) Conceptual visualization of LWFA simulation. }
  \label{fig:lwfa_performence}
\end{figure}

In the more complex Laser-Wakefield Acceleration (LWFA) workload, \textbf{MatrixPIC} achieves a significant speedup of up to \textbf{2.62$\times$} (Figure~\ref{fig:lwfa_performence}) in total simulation time. The underlying reason for this strong performance is twofold. Firstly, the shock front inherent in LWFA leads to high-density particle regions, which are highly amenable to \textbf{MatrixPIC}'s vectorization. Secondly, the potential negative impact of substantial particle movement is effectively mitigated by our $O(1)$ amortized incremental sorting algorithm and adaptive re-sorting policy, which efficiently manage data locality.

It is noteworthy, however, that this performance advantage is density-dependent. In the LWFA scenario, we observe that at particle densities below 8 PPC, the performance of \textbf{MatrixPIC} can fall below that of the baseline. This behavior is consistent with our findings in the uniform plasma experiments and is attributed to the challenge of leveraging high-density compute hardware in sparse regions, where framework overheads are not fully amortized. For realistic LWFA simulations, this is less of a concern, as the scientifically critical regions—such as the plasma wake—typically feature high particle densities (32 to 256 PPC). This range aligns perfectly with the operational sweet spot for \textbf{MatrixPIC}. Consequently, for production environments, we recommend a hybrid execution strategy where a fallback to a scalar or optimized VPU kernel is employed in regions where particle density drops below a threshold (e.g., 8 PPC) to guarantee the best overall performance.

\subsection{Ablation Study}
% \subsection{Component-wise Performance}
\label{subsec:ablation_study}

\begin{figure}[t]
\centering
\includegraphics[width=\linewidth]{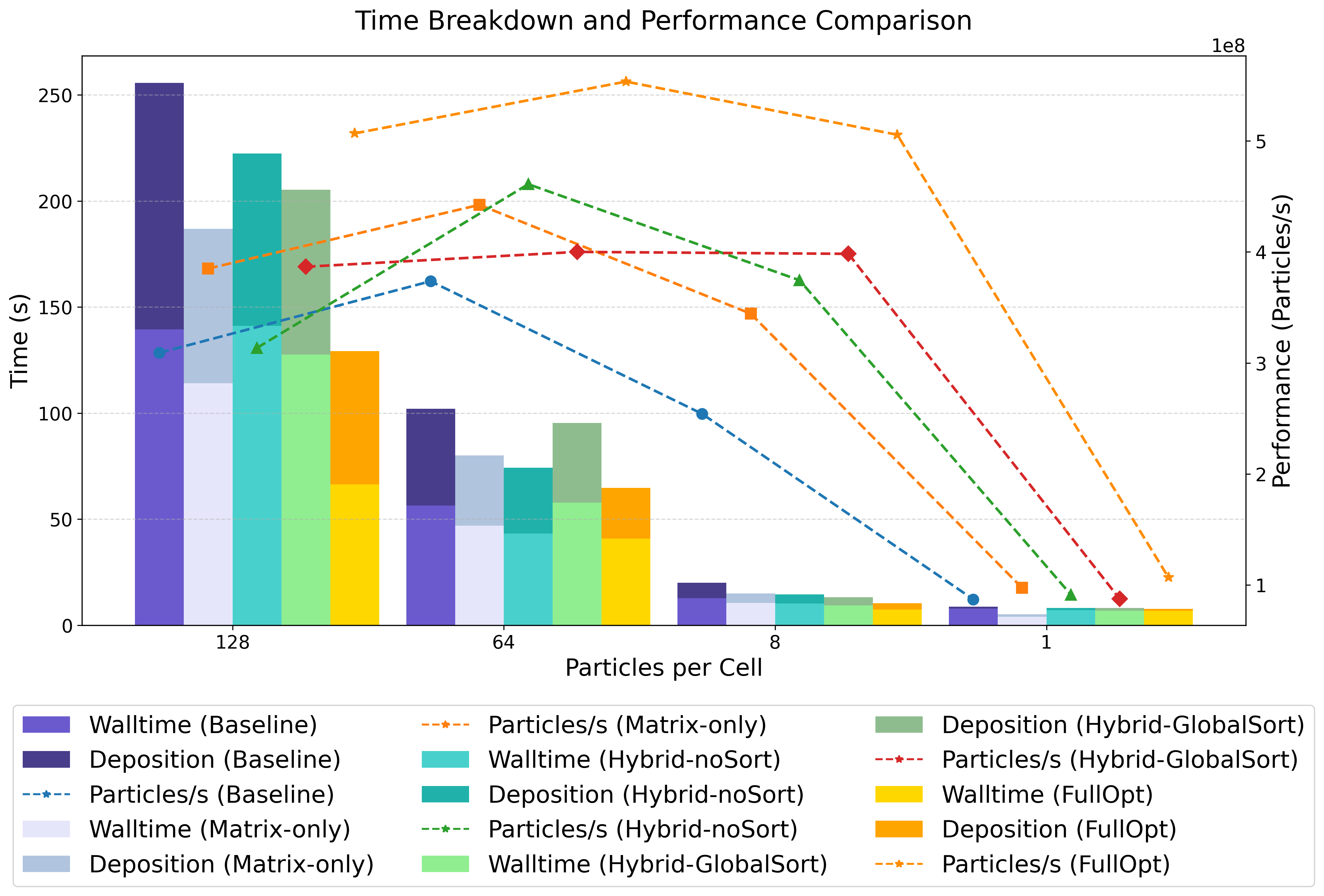}
\caption{Ablation study results showing the performance of different \textbf{MatrixPIC} configurations compared to the baseline \texttt{WarpX}.}
\label{fig:ablation_study_plots}
\end{figure}

To understand the impact of our key optimization components, we first conducted a high-level ablation study comparing our intermediate designs against the baseline. Figure~\ref{fig:ablation_study_plots} illustrates the total wall time and particle throughput across varying particle densities (PPC) for the uniform plasma workload.

The analysis reveals key trade-offs among the intermediate configurations. The \textbf{Matrix-only} design, leveraging the MPU's raw computational power, achieves the shortest wall time at high density (PPC=128) with \textasciitilde187~s; here, its kernel time (\textasciitilde72.95~s) is substantially lower than the baseline's (\textasciitilde116~s). The \textbf{Hybrid-noSort} approach performs exceptionally well at medium density (PPC=64), achieving the lowest overall wall time of \textasciitilde81.4~s and the highest particle throughput of \textasciitilde4.6$\times10^8$ P/s. However, VPU-MPU interaction overheads cause its performance to degrade at higher densities, where its wall time at PPC=128 (\textasciitilde237.11~s) is significantly longer than the \textbf{Matrix-only} version. Conversely, the \textbf{Hybrid-GlobalSort} configuration is frequently bottlenecked by the high cost of non-incremental sorting, evident from its long wall time at PPC=128 (\textasciitilde219.2~s).

Despite the varying performance of these intermediate steps, the fully-integrated \textbf{MatrixPIC} design consistently delivers the best overall wall time and highest throughput across the tested scenarios, validating the effectiveness of our complete, hybrid approach.

\subsection{Comparative Analysis against VPU Baselines}
% \subsection{Detailed Kernel Breakdown and Architectural Insights}

To quantify the benefits of our innovations more directly and to rigorously compare \textbf{MatrixPIC} against state-of-the-art VPU methods, we conducted a second, more detailed set of comparative studies. The core deposition kernel's performance was measured with high-precision inline assembly counters to isolate its execution time.

\begin{table}[h]
    \centering
    \small
    \setlength{\tabcolsep}{3pt} % Reduce inter-column spacing
    \caption{Performance Breakdown of the First-Order (CIC) Deposition Kernel.}
    \label{tab:cic_breakdown}
    \begin{tabular}{lcccc}
        \toprule
        \textbf{Configuration} & \textbf{Total} & \textbf{Preproc.} & \textbf{Compute} & \textbf{Sort}  \\
        & \textbf{(s)} & \textbf{(s)} & \textbf{(s)} & \textbf{(s)} \\
        \midrule
        Baseline (WarpX)        & 74.13 & 17.39 & 56.74 & --   \\
        Baseline+IncrSort       & 45.64 & 20.74 & 19.71 & 5.19 \\
        Rhocell (auto-vec)      & 54.89 & 19.89 & 34.75 & --   \\
        Rhocell+IncrSort        & 44.81 & 20.49 & 23.38 & 4.63 \\
        Rhocell+IncrSort (VPU)  & 34.13 & 7.66  & 21.04 & 5.11 \\
        \textbf{MatrixPIC (FullOpt)} & \textbf{24.90} & \textbf{5.33}  & \textbf{15.10} & \textbf{4.31} \\
        \bottomrule
    \end{tabular}
\end{table}

\paragraph{First-Order (CIC) Scheme Analysis}
Table~\ref{tab:cic_breakdown} presents the performance breakdown for the CIC kernel. The measurements confirm that our fully-optimized \textbf{MatrixPIC} kernel achieves a \textbf{2.98x} speedup over the compiler auto-vectorized \texttt{WarpX} baseline.

The standalone impact of our incremental sorting algorithm is significant, delivering a 1.62$\times$ speedup when added to the baseline. This locality-enhancing strategy is even 1.2$\times$ faster than the \texttt{Rhocell} baseline, suggesting that for this problem, improving data access patterns can be more impactful than solely restructuring data to mitigate atomic conflicts. Crucially, the auto-vectorized \texttt{Rhocell} is limited by two factors: its preprocessing stages, which compilers struggle to vectorize efficiently, and its unordered particle access, which leads to weaker compute performance compared to the sorted approaches (as shown in Table~\ref{tab:cic_breakdown}). Indeed, adding our incremental sorter to the \texttt{Rhocell} baseline significantly boosts its computation phase by 1.49$\times$. This motivates our creation of the strongest possible VPU baseline, \texttt{Rhocell+IncrSort (VPU)}, which is fully hand-tuned with VPU intrinsics. Even against this highly-optimized competitor, \textbf{MatrixPIC} is still \textbf{1.37$\times$} faster. This result demonstrates both the MPU's architectural superiority and the overall effectiveness of our co-design, especially given that the CIC scheme utilizes only 25\% of the MPU tile.

\begin{table}[h]
    \centering
    \small
    \setlength{\tabcolsep}{4pt} % Reduce inter-column spacing
    \caption{Performance Breakdown of the Third-Order (QSP) Deposition Kernel.}
    \label{tab:qsp_breakdown}
    \begin{tabular}{lcccc}
        \toprule
        \textbf{Configuration} & \textbf{Total} & \textbf{Preproc.} & \textbf{Compute} & \textbf{Sort} \\
        & \textbf{(s)} & \textbf{(s)} & \textbf{(s)} & \textbf{(s)} \\
        \midrule
        Baseline (WarpX)       & 12.19 & 0.38 & 11.82 & -- \\
        Baseline+IncrSort      & 3.44  & 0.39 & 3.02  & 0.03 \\
        Rhocell+IncrSort (VPU) & 2.81  & 0.13 & 2.63  & 0.04 \\
        \textbf{MatrixPIC (FullOpt)} & \textbf{1.39}  & \textbf{0.13} & \textbf{1.22}  & \textbf{0.03} \\
        \bottomrule
    \end{tabular}
\end{table}

\paragraph{Third-Order (QSP) Scheme Analysis}
The performance advantage of \textbf{MatrixPIC} grows substantially with the more computationally dense, third-order QSP kernel. This is because its higher arithmetic intensity increases MPU tile utilization from 25\% (for the CIC scheme) to 50\%. As shown in Table~\ref{tab:qsp_breakdown}, this results in a remarkable \textbf{8.7$\times$} speedup over the baseline and \textbf{2.0$\times$} over the best hand-tuned VPU implementation. This demonstrates that the QSP scheme's higher computational density effectively amortizes the overheads of sorting and data preprocessing. The performance breakdown confirms this: the 2$\times$ speedup over the VPU baseline is driven almost entirely by the computation phase, while the relative cost of sorting drops to just 2.2\% of the total kernel time (down from 11.3\% for CIC). This trend strongly suggests that our framework is particularly well-suited for higher-order scientific computations, where the architectural advantages of high-density hardware like MPUs can be fully realized.

\subsection{Cross-Platform Efficiency}

To situate our results in a broader architectural context, we analyze the cross-platform efficiency in Table~\ref{tab:peak_perf}. A direct adaptation of the baseline kernel to the LX2 CPU achieves a mere 9.84\% of theoretical peak FP64 performance. This is only 33.1\% of the efficiency attained by its CUDA counterpart on an NVIDIA A800 GPU platform (29.76\%), highlighting the formidable challenge posed by the GPU's massively parallel architecture for traditional CPU execution.

\begin{table}[htb]
\centering
\small % Use smaller font for compactness
\setlength{\tabcolsep}{3pt} % Reduce inter-column spacing
\caption{Cross-platform kernel efficiency (\% of theoretical peak FP64). Reference data from the WarpX manuscript~\cite{WarpX2022} reports the overall application efficiency (including gather and deposition) on whole systems and is provided for context.}
\label{tab:peak_perf}
\begin{tabular}{llc}
    \toprule
    \textbf{System} & \textbf{Config.} & \textbf{Peak Efficiency (\%)} \\
    \midrule
    \multicolumn{3}{l}{\textit{Our Results on QSP Kernel}} \\
        \textbf{LX2 CPU} & \textbf{MatrixPIC (Ours)} & \textbf{83.08} \\
        LX2 CPU & Rhocell+IncrSort (VPU) & 54.58 \\
        LX2 CPU & Baseline & 9.84 \\
    \midrule
        \textit{\textbf{NVIDIA A800}} & \textit{\textbf{Baseline (CUDA)}} & \textit{\textbf{29.76}} \\
    \midrule
    \multicolumn{3}{l}{\textit{Reference Data (Overall App. Efficiency)}} \\
        Perlmutter (A100) & Baseline (App.) & 12.9 \\
        Summit (V100) & Baseline (App.) & 8.3 \\
        Frontier (MI250x) & Baseline (App.) & 3.3 \\
        Fugaku (A64FX) & Baseline (App.) & 1.1 \\
    \bottomrule
\end{tabular}
\end{table}
% \vspace{-6pt}

In stark contrast, our final \textbf{MatrixPIC} implementation elevates the efficiency to an impressive \textbf{83.08\%}. This dramatic improvement is attributed to two primary factors. First, the enhanced data locality from our optimized data layout and sorting is fundamental; this alone boosts the hand-tuned VPU implementation to a strong \textbf{54.58\%} efficiency. Second, building on this foundation of data locality, \textbf{MatrixPIC} fully utilizes the MPU's specialized compute capabilities to push performance even further.

Ultimately, our results demonstrate the superiority of a methodology that deeply exploits platform-specific architectural features. By mapping the algorithm to the MPU's native computational paradigm, our framework fully extracts the performance potential from the specialized hardware. This finding validates that for compute patterns amenable to matrix-based execution, emerging CPU-integrated accelerators can offer a more potent and efficient solution than general-purpose GPU cores, unlocking new avenues for performance in particle-based simulations.

\section{Conclusion and Future Work}
\label{sec:Conclusion}

In this work, we introduced \textbf{MatrixPIC}, a novel framework that leverages Matrix Processing Units (MPUs) to accelerate the deposition phase of particle-mesh algorithms, highlighting its general applicability to scientific computing. By holistically co-designing a hybrid VPU-MPU kernel, an optimized \texttt{rhocell} data structure, and a low-overhead incremental sorting algorithm, our framework achieves significant performance gains over the highly-optimized \texttt{WarpX} baseline. Evaluations demonstrate its effectiveness in first-order CIC scenarios, yielding speedups of up to \textbf{1.19$\times$} in uniform plasma simulations and \textbf{2.63$\times$} in realistic Laser-Wakefield Acceleration workloads.

Ablation studies confirm the impact of our individual contributions, with the incremental sorting algorithm alone delivering a \textbf{1.62$\times$} kernel speedup. However, the framework's full potential is unlocked with higher-order algorithms. For the third-order QSP scheme, \textbf{MatrixPIC} achieves a remarkable \textbf{8.7$\times$ speedup} over the baseline and is \textbf{2.0$\times$ faster} than the best hand-tuned VPU implementation. This result strongly validates our central thesis: co-designing for high-density hardware is crucial for future scientific algorithms that demand higher precision and complexity.

Furthermore, our cross-platform efficiency analysis reveals the profound impact of our co-design approach on hardware utilization. MatrixPIC on the MPU-enabled CPU achieves a remarkable \textbf{83.08\%} of theoretical peak performance, effectively saturating the hardware's capabilities. This level of efficiency is nearly \textbf{2.8$\times$} higher than that of the highly-optimized CUDA implementation on a data center GPU.This finding validates that CPU-integrated matrix hardware, when paired with algorithms reformulated for the MOPA paradigm, is a potent and efficient platform for this class of scientific simulations.

\paragraph{Limitations and Future Work}
Despite these advances, our work has several limitations. The implementation is confined to direct deposition, lacking support for charge-conserving schemes (e.g., Esirkepov~\cite{Esirkepov2001Exact,Morse1971Weibel} or Vay~\cite{Vay2013Domain}), and validating its generality through case studies in other domains, such as the Particle-Mesh (PM) method in N-body cosmology simulations ~\cite{couchman1991mesh, breton2025pysco}, remains a key area for future work. Moreover, the optimizations are platform-specific and target only the deposition phase. Future work will focus on extending \textbf{MatrixPIC} to support these advanced schemes and optimizing the gather phase. We also plan to implement adaptive strategies, such as a fallback to VPU or scalar kernels in low-density regions~\cite{Beck2019AdaptiveSIMD}, and to explore porting the framework to other MPU-enabled architectures to broaden its applicability.

\begin{acks}
We sincerely thank our shepherd Kapil Vaswani and all the anonymous reviewers for their valuable feedback. We also thank Zhenyu Wang of the Institute of Plasma Physics, Chinese Academy of Sciences, for insightful discussions and guidance during our experiments. This research was supported by Guangdong S\&T Program under Grant No. 2024B0101040005, the National Natural Science Foundation of China (NSFC): No.62461146204 and No.62502552, Guangdong Province Special Support Program for Cultivating High-Level Talents: 2021TQ06X160.

This research used the open-source particle-in-cell code \href{https://github.com/BLAST-WarpX/warpx}{WarpX}.
We acknowledge all WarpX contributors.
\end{acks}

\bibliographystyle{ACM-Reference-Format}
\bibliography{sample-base}

@String{Computing = "Computing" }

@String{Computer = "{IEEE} Computer" }

@String{Springer = "Springer-Verlag" }

@article{SMILEI2018,
  author       = {Derouillat, Jo{\"e}l and Beck, Antoine and P{\'e}rez, Fr{\'e}d{\'e}ric and Vinci, Thomas and Chiaramello, Micka{\"e}l and Grassi, A. and Flentje, M. and Bouchard, G. and Plotnikov, I. and Aunai, N. and Dargent, J. and Riconda, C. and Grech, M.},
  year         = {2018},
  title        = {{SMILEI}: a collaborative, open-source, multi-purpose particle-in-cell code for plasma simulation},
  journal      = {Computer Physics Communications},
  volume       = {222},
  pages        = {351--373},
  doi          = {10.1016/j.cpc.2017.09.024},
  url          = {https://doi.org/10.1016/j.cpc.2017.09.024},
  month        = {jan}
}

@inproceedings{WarpX2022,
author={Fedeli, Luca and Huebl, Axel and Boillod-Cerneux, France and Clark, Thomas and Gott, Kevin and Hillairet, Conrad and Jaure, Stephan and Leblanc, Adrien and Lehe, Rémi and Myers, Andrew and Piechurski, Christelle and Sato, Mitsuhisa and Zaim, Neïl and Zhang, Weiqun and Vay, Jean-Luc and Vincenti, Henri},
  booktitle={SC22: International Conference for High Performance Computing, Networking, Storage and Analysis}, 
  title={Pushing the Frontier in the Design of Laser-Based Electron Accelerators with Groundbreaking Mesh-Refined Particle-In-Cell Simulations on Exascale-Class Supercomputers}, 
  year={2022},
  volume={},
  number={},
  pages={1-12},
  doi={10.1109/SC41404.2022.00008}
}

@inproceedings{Barsamian2018StrictBinning,
  author       = {Barsamian, Yann and Chargu{\'e}raud, Arthur and Hirstoaga, Sever A. and Mehrenberger, Michel},
  year         = {2018},
  title        = {Efficient Strict-Binning Particle-in-Cell Algorithm for Multi-core {SIMD} Processors},
  booktitle    = {Euro-Par 2018: Parallel Processing},
  publisher    = {Springer},
  series       = {Lecture Notes in Computer Science},
  volume       = {11014},
  pages        = {749--763},
  doi          = {10.1007/978-3-319-96983-1_53},
  url          = {https://doi.org/10.1007/978-3-319-96983-1_53},
  month        = {aug}
}

@article{Tajima1982Laser,
  author  = {T. Tajima and J.~M. Dawson},
  title   = {Laser accelerator by plasma waves},
  journal = {AIP Conference Proceedings},
  volume  = {91},
  number  = {1},
  pages   = {69--93},
  year    = {1982},
  month   = sep,
  doi     = {10.1063/1.33805},
  url     = {https://doi.org/10.1063/1.33805}
}

@article{Esarey1996Overview,
 author={Esarey, E. and Sprangle, P. and Krall, J. and Ting, A.},
  journal={IEEE Transactions on Plasma Science}, 
  title={Overview of plasma-based accelerator concepts}, 
  year={1996},
  volume={24},
  number={2},
  pages={252-288},
  doi={10.1109/27.509991}
}

@inproceedings{Li2021YHPIC,
  author       = {Li, Biao and Zhang, Qingyang and Liu, Jie and Chen, Xinhai and Zhu, Xiaoxiong and Wang, Qinglin and Zhuo, Hongbin},
  year         = {2021},
  title        = {Kinetic Simulations of Laser Plasma Interaction with 42.9 Trillion Particles and 10.4 Billion Grids},
  booktitle    = {Proceedings of HPC China 2021}, 
}

@article{Weidman2021SMEArm,
  author       = {Weidman, Martin},
  year         = {2021},
  title        = {Introducing the Scalable Matrix Extension for the {Armv9-A} architecture},
  howpublished = {Arm Community Blog},
  month        = {aug}, 
  url          = {https://community.arm.com/arm-community-blogs/b/architectures-and-processors-blog/posts/scalable-matrix-extension-armv9-a-architecture},
  note         = {Accessed May 2024}
}

@article{Bowers2008VPIC,
  author    = {Bowers, K. J. and Albright, B. J. and Yin, L. and Bergen, B. and Kwan, T. J. T.},
  year      = {2008},
  title     = {Ultrahigh performance three-dimensional electromagnetic relativistic kinetic plasma simulation},
  journal   = {Physics of Plasmas},
  volume    = {15},
  number    = {5},
  pages     = {055703},
  doi       = {10.1063/1.2840133},
  url       = {https://doi.org/10.1063/1.2840133},
  month     = {may}
}

@article{Bird2022VPIC2,
  author    = {Bird, R. and Tan, N. and Luedtke, S. V. and Harrell, S. L. and Taufer, M. and Albright, B.},
  year      = {2022},
  title     = {{VPIC 2.0}: Next Generation Particle-in-Cell Simulations},
  journal   = {IEEE Transactions on Parallel and Distributed Systems},
  volume    = {33},
  number    = {4},
  pages     = {952--963},
  doi       = {10.1109/TPDS.2021.3084795},
  url       = {https://doi.org/10.1109/TPDS.2021.3084795},
  month     = {apr}
}

@misc{Chandrasekaran2019PIConGPUSummit,
  title        = {Oak Ridge National Laboratory Case Study: ORNL Expands Possibilities for Plasma Physics with Open and Portable AMD ROCm},
  author       = {Advanced Micro Devices, Inc.},
  year         = {2020},
  howpublished = {\url{https://www.amd.com/content/dam/amd/en/documents/resources/case-studies/oak-ridge-national-laboratory-case-study.pdf}},
  note         = {[Online; accessed 14 May 2025]}
}

@article{Itai1981SparseTables,
  title={A sparse table implementation of priority queues},
  author={Itai, Alon and Konheim, Alan G and Rodeh, Michael},
  booktitle={International Colloquium on Automata, Languages, and Programming},
  pages={417--431},
  year={1981},
  organization={Springer}
}

@inproceedings{Bender2000CacheObliviousBtrees,
  author={Bender, M.A. and Demaine, E.D. and Farach-Colton, M.},
  booktitle={Proceedings 41st Annual Symposium on Foundations of Computer Science}, 
  title={Cache-oblivious B-trees}, 
  year={2000},
  volume={},
  number={},
  pages={399-409},
  doi={10.1109/SFCS.2000.892128}
}

@misc{bib_apple_com,
  author       = {{Apple Inc.}},
  title        = {Apple introduces M4 chip},
  howpublished = {\url{https://www.apple.com/hk/en/newsroom/2024/05/apple-introduces-m4-chip/}},
  month        = may,
  day          = {07},
  year         = {2024},
  note         = {[Online; accessed 14 May 2025]}
}

@techreport{bib_AMD_MI300X,
  author       = {{Advanced Micro Devices, Inc.}},
  title        = {{AMD Instinct™ MI300X Accelerator Data Sheet}},
  institution  = {{Advanced Micro Devices, Inc.}},
  year         = {2023},
  month        = may,
  howpublished = {\url{https://www.amd.com/content/dam/amd/en/documents/instinct-tech-docs/data-sheets/amd-instinct-mi300x-data-sheet.pdf}},
  note         = {[Online; accessed 14 May 2025]}
}

@article{bib_CFL_Condi_placeholder,
  author    = {R. E. Groenewald and A. Veksler and F. Ceccherini and A. Necas and B. S. Nicks and D. C. Barnes and T. Tajima and S. A. Dettrick},
  title     = {Accelerated kinetic model for global macro stability studies of high-beta fusion reactors},
  journal   = {Physics of Plasmas},
  volume    = {30},
  number    = {12},
  pages     = {122508},
  year      = {2023},
  month     = dec,
  doi       = {10.1063/5.0178288}
}

@article{Vincenti2017SIMD,
  author       = {Vincenti, H{\'e}l{\`e}ne and Lobet, Maxime and Lehe, R{\'e}mi and Sasanka, Rohan and Vay, Jean-Luc},
  year         = {2017},
  title        = {An efficient and portable {SIMD} algorithm for charge/current deposition in Particle-In-Cell codes},
  journal      = {Computer Physics Communications},
  volume       = {210},
  pages        = {145--154},
  doi          = {10.1016/j.cpc.2016.08.023},
  url          = {https://doi.org/10.1016/j.cpc.2016.08.023},
  month        = {jan}
}

@inproceedings{Barsamian2017DataStructures,
  author       = {Barsamian, Yann and Violard, {\'E}ric and Hirstoaga, Sever A.},
  year         = {2017},
  title        = {Efficient Data Structures for a Hybrid Parallel and Vectorized Particle-in-Cell Code},
  booktitle    = {2017 IEEE International Parallel and Distributed Processing Symposium Workshops (IPDPSW)},
  pages        = {1168--1177},
  publisher    = {IEEE},
  address      = {Orlando, FL, USA},
  doi          = {10.1109/IPDPSW.2017.74},
  url          = {https://doi.org/10.1109/IPDPSW.2017.74},
  month        = {may}
}

@article{couchman1991mesh,
  title={Mesh-refined P3M-A fast adaptive N-body algorithm},
  author={Couchman, HMP},
  journal={Astrophysical Journal, Part 2-Letters (ISSN 0004-637X), vol. 368, Feb. 20, 1991, p. L23-L26. Research supported by NSERC.},
  volume={368},
  pages={L23--L26},
  year={1991}
}

@article{cheatham1995molecular,
  title={Molecular dynamics simulations on solvated biomolecular systems: the particle mesh Ewald method leads to stable trajectories of DNA, RNA, and proteins},
  author={Cheatham, TE III and Miller, JL and Fox, T and Darden, TA and Kollman, PA},
  journal={Journal of the American Chemical Society},
  volume={117},
  number={14},
  pages={4193--4194},
  year={1995},
  publisher={ACS Publications}
}

@inproceedings{frigo2009reducers,
  title={Reducers and other Cilk++ hyperobjects},
  author={Frigo, Matteo and Halpern, Pablo and Leiserson, Charles E and Lewin-Berlin, Stephen},
  booktitle={Proceedings of the twenty-first annual symposium on Parallelism in algorithms and architectures},
  pages={79--90},
  year={2009}
}

@inproceedings{lee2015efficiently,
  title={Efficiently detecting races in cilk programs that use reducer hyperobjects},
  author={Lee, I-Ting Angelina and Schardl, Tao B},
  booktitle={Proceedings of the 27th ACM Symposium on Parallelism in Algorithms and Architectures},
  pages={111--122},
  year={2015}
}

@inproceedings{simakov2024first,
    author = {Simakov, Nikolay A. and Jones, Matthew D. and Furlani, Thomas R. and Siegmann, Eva and Harrison, Robert J.},
    title = {First Impressions of the NVIDIA Grace CPU Superchip and NVIDIA Grace Hopper Superchip for Scientific Workloads},
    year = {2024},
    isbn = {9798400716522},
    publisher = {Association for Computing Machinery},
    address = {New York, NY, USA},
    url = {https://doi.org/10.1145/3636480.3637097},
    doi = {10.1145/3636480.3637097},
    booktitle = {Proceedings of the International Conference on High Performance Computing in Asia-Pacific Region Workshops},
    pages = {36–44},
    numpages = {9},
    location = {Nagoya, Japan},
    series = {HPCAsia '24 Workshops}
}

@article{abraham2011optimization,
  title={Optimization of parameters for molecular dynamics simulation using smooth particle-mesh Ewald in GROMACS 4.5},
  author={Abraham, Mark J and Gready, Jill E},
  journal={Journal of computational chemistry},
  volume={32},
  number={9},
  pages={2031--2040},
  year={2011},
  publisher={Wiley Online Library}
}

@article{breton2025pysco,
  title={PySCo: A fast particle-mesh N-body code for modified gravity simulations in Python},
  author={Breton, Michel-Andr{\`e}s},
  journal={Astronomy \& Astrophysics},
  volume={695},
  pages={A170},
  year={2025},
  publisher={EDP Sciences}
}

@article{Barsamian2018DataLayouts,
  author       = {Barsamian, Yann and Hirstoaga, Sever A. and Violard, {\'E}ric},
  year         = {2018},
  title        = {Efficient data layouts for a three-dimensional electrostatic Particle-in-Cell code},
  journal      = {Journal of Computational Science},
  volume       = {27},
  pages        = {345--356},
  doi          = {10.1016/j.jocs.2018.06.004},
  url          = {https://doi.org/10.1016/j.jocs.2018.06.004},
  month        = {jul}
}

@inproceedings{Zhao2024StencilSME,
  author       = {Zhao, Wenxuan and Yuan, Liang and Yan, Baicheng and Ma, Penghao and Zhang, Yunquan and Wang, Long and Wang, Zhe},
  year         = {2024},
  title        = {Stencil Computation with Vector Outer Product},
  booktitle    = {Proceedings of the 38th ACM International Conference on Supercomputing (ICS '24)},
  pages        = {247--258},
  publisher    = {ACM},
  address      = {Kyoto, Japan},
  doi          = {10.1145/3650200.3656611},
  url          = {https://doi.org/10.1145/3650200.3656611},
  month        = {jun}
}

@article{Remke2024HelloSME,
 author={Remke, Stefan and Breuer, Alexander},
  booktitle={SC24-W: Workshops of the International Conference for High Performance Computing, Networking, Storage and Analysis}, 
  title={Hello SME! Generating Fast Matrix Multiplication Kernels Using the Scalable Matrix Extension}, 
  year={2024},
  volume={},
  number={},
  pages={1443-1454},
  doi={10.1109/SCW63240.2024.00185}
}

@techreport{HarrisonA64FXSVE,
  author       = {Harrison, Robert J.},
  year         = {2020},
  title        = {Performance engineering on {A64FX} with {SVE} intrinsics (Early experience on Ookami)},
  institution  = {Stony Brook University, Institute for Advanced Computational Science},
  type         = {Presentation},
  url          = {https://www.stonybrook.edu/commcms/ookami/support/_docs/RJHACMCF21.pdf}, 
  note         = {Accessed May 2024}
}

@article{Bender2007AdaptivePMA,
    author = {Bender, Michael A. and Hu, Haodong},
    title = {An adaptive packed-memory array},
    year = {2007},
    issue_date = {November 2007},
    publisher = {Association for Computing Machinery},
    address = {New York, NY, USA},
    volume = {32},
    number = {4},
    issn = {0362-5915},
    url = {https://doi.org/10.1145/1292609.1292616},
    doi = {10.1145/1292609.1292616},
    month = nov,
    pages = {26–es},
    numpages = {43}
}

@article{Bowers2001HybridSort,
    title = {Accelerating a Particle-in-Cell Simulation Using a Hybrid Counting Sort},
    journal = {Journal of Computational Physics},
    volume = {173},
    number = {2},
    pages = {393-411},
    year = {2001},
    issn = {0021-9991},
    doi = {https://doi.org/10.1006/jcph.2001.6851},
    url = {https://www.sciencedirect.com/science/article/pii/S0021999101968512},
    author = {K.J Bowers}
}

@inproceedings{Bussmann2013PIConGPU,
    author = {Bussmann, M. and Burau, H. and Cowan, T. E. and Debus, A. and Huebl, A. and Juckeland, G. and Kluge, T. and Nagel, W. E. and Pausch, R. and Schmitt, F. and Schramm, U. and Schuchart, J. and Widera, R.},
    title = {Radiative signatures of the relativistic Kelvin-Helmholtz instability},
    year = {2013},
    isbn = {9781450323789},
    publisher = {Association for Computing Machinery},
    address = {New York, NY, USA},
    url = {https://doi.org/10.1145/2503210.2504564},
    doi = {10.1145/2503210.2504564},
    booktitle = {Proceedings of the International Conference on High Performance Computing, Networking, Storage and Analysis},
    articleno = {5},
    numpages = {12},
    location = {Denver, Colorado},
    series = {SC '13}
}

@inproceedings{Zenker2016PIConGPUOpenPower,
  author       = {Zenker, E. and Bussmann, M. and Juckeland, G. and Debus, A. and Huebl, A. and Widera, R. and Kluge, T. and Schramm, U.},
  year         = {2016},
  title        = {Performance-Portable Many-Core Plasma Simulations: Porting {PIConGPU} to {OpenPOWER} and Beyond},
  booktitle    = {High Performance Computing. ISC High Performance 2016},
  editor       = {Taufer, Michela and Mohr, Bernd and Kunkel, Julian M.},
  publisher    = {Springer},
  series       = {Lecture Notes in Computer Science},
  volume       = {9945},
  pages        = {293--301},
  doi          = {10.1007/978-3-319-46079-6_21},
  url          = {https://doi.org/10.1007/978-3-319-46079-6_21},
  month        = {jun} 
}

@article{Decyk2014PICEmergingArch,
  author       = {Decyk, Viktor K. and Singh, Tajendra V.},
  year         = {2014},
  title        = {Particle-in-Cell algorithms for emerging computer architectures},
  journal      = {Computer Physics Communications},
  volume       = {185},
  number       = {3},
  pages        = {708--719},
  doi          = {10.1016/j.cpc.2013.10.013},
  url          = {https://doi.org/10.1016/j.cpc.2013.10.013},
  month        = {mar}
}

@inproceedings{Durand2012PMA_MovingParticles,
  TITLE = {{A Packed Memory Array to Keep Moving Particles Sorted}},
  AUTHOR = {Durand, Marie and Raffin, Bruno and Faure, Fran{\c c}ois},
  URL = {https://inria.hal.science/hal-00762593},
  BOOKTITLE = {{Ninth Workshop on Virtual Reality Interaction and Physical Simulation (VRIPHYS)}},
  ADDRESS = {Darmstadt, Germany},
  EDITOR = {Jan Bender and Arjan Kuijper and Dieter W. Fellner and Eric Gu{\'e}rin},
  PUBLISHER = {{The Eurographics Association}},
  SERIES = {Ninth Workshop on Virtual Reality Interaction and Physical Simulation (VRIPHYS)},
  PAGES = {69-77},
  YEAR = {2012},
  MONTH = Dec,
  HAL_ID = {hal-00762593},
  HAL_VERSION = {v1},
}

@inproceedings{WarpXOptaneIPDPSW2021,
    author = {Ren, Jie and Luo, Jiaolin and Peng, Ivy and Wu, Kai and Li, Dong},
    title = {Optimizing large-scale plasma simulations on persistent memory-based heterogeneous memory with effective data placement across memory hierarchy},
    year = {2021},
    isbn = {9781450383356},
    publisher = {Association for Computing Machinery},
    address = {New York, NY, USA},
    url = {https://doi.org/10.1145/3447818.3460356},
    doi = {10.1145/3447818.3460356},
    booktitle = {Proceedings of the 35th ACM International Conference on Supercomputing},
    pages = {203–214},
    numpages = {12},
    location = {Virtual Event, USA},
    series = {ICS '21}
}

@book{Birdsall1985PlasmaPhysics,
  title={Plasma physics via computer simulation},
  author={Birdsall, Charles K and Langdon, A Bruce},
  year={2018},
  publisher={CRC press}
}

@book{Hockney1988ComputerSim,
  author    = {Hockney, Roger W. and Eastwood, James W.},
  title     = {Computer Simulation Using Particles},
  publisher = {Taylor \& Francis}, 
  year      = {1988}
}

@inproceedings{Nakashima2017PIC,
  author       = {Nakashima, Hiroshi and Summura, Yoshiki and Kikura, Keisuke and Miyake, Yohei},
  year         = {2017},
  title        = {Large Scale Manycore-Aware {PIC} Simulation with Efficient Particle Binning},
  booktitle    = {2017 IEEE International Parallel and Distributed Processing Symposium (IPDPS)},
  pages        = {202--211},
  publisher    = {IEEE},
  address      = {Orlando, FL, USA},
  doi          = {10.1109/IPDPS.2017.65},
  url          = {https://doi.org/10.1109/IPDPS.2017.65},
  month        = {may}
}

@article{Beck2019AdaptiveSIMD,
  author       = {Beck, Antoine and Derouillat, Jo{\"e}l and Lobet, Maxime and Farjallah, Anouar and Massimo, Francesco and Zemzemi, Ikbel and P{\'e}rez, Fr{\'e}d{\'e}ric and Vinci, Thomas and Grech, Micka{\"e}l},
  year         = {2019},
  title        = {Adaptive SIMD optimizations in particle-in-cell codes with fine-grain particle sorting},
  journal      = {Computer Physics Communications},
  volume       = {244},
  pages        = {246--263},
  doi          = {10.1016/j.cpc.2019.05.001},
  url          = {https://doi.org/10.1016/j.cpc.2019.05.001},
  month        = {nov}
}

@book{bib_CFL_condition_PIC_ref,
    author = {Moura, Carlos A. de and Kubrusly, Carlos S.},
    title = {The Courant-Friedrichs-Lewy (CFL) Condition: 80 Years After Its Discovery},
    year = {2012},
    isbn = {0817683933},
    publisher = {Birkh\"{a}user Basel}
}

@article{Esirkepov2001Exact,
  author  = {T.~Zh. Esirkepov},
  title   = {Exact Charge Conservation Scheme for Particle-In-Cell Simulation with an Arbitrary Form-Factor},
  journal = {Computer Physics Communications},
  volume  = {135},
  number  = {2},
  pages   = {144--153},
  year    = {2001},
  month   = apr,
  doi     = {10.1016/S0010-4655(00)00228-9},
  url     = {https://doi.org/10.1016/S0010-4655(00)00228-9}
}

@article{Morse1971Weibel,
    author = {Morse, R. L. and Nielson, C. W.},
    title = {Numerical Simulation of the Weibel Instability in One and Two Dimensions},
    journal = {The Physics of Fluids},
    volume = {14},
    number = {4},
    pages = {830-840},
    year = {1971},
    month = {04},
    issn = {0031-9171},
    doi = {10.1063/1.1693518},
    url = {https://doi.org/10.1063/1.1693518},
    eprint = {https://pubs.aip.org/aip/pfl/article-pdf/14/4/830/12783644/830\_1\_online.pdf}
}

@article{Vay2013Domain,
    title = {A domain decomposition method for pseudo-spectral electromagnetic simulations of plasmas},
    journal = {Journal of Computational Physics},
    volume = {243},
    pages = {260-268},
    year = {2013},
    issn = {0021-9991},
    doi = {https://doi.org/10.1016/j.jcp.2013.03.010},
    url = {https://www.sciencedirect.com/science/article/pii/S0021999113001873},
    author = {Jean-Luc Vay and Irving Haber and Brendan B. Godfrey}
}

\onecolumn
\appendix

\section{Appendix: Parameter Table}
\label{appen:parameter}

\begin{table*}[ht]
\centering
\caption{Key Parameters for the Uniform Plasma and LWFA Simulation Workloads.}
\label{tab:workload_parameters}
\resizebox{\textwidth}{!}{%
\begin{tabular}{lcc}
\toprule
\textbf{Parameter} & \textbf{Uniform Plasma} & \textbf{Laser-Wakefield Acceleration (LWFA)} \\
\midrule
\textbf{Primary Goal} & Assess SIMD kernel efficiency & Evaluate performance in dynamic, real-world scenario \\
\textbf{Domain Setup} & & \\
\quad \texttt{geometry.dims} & 3 & 3 \\
\quad \texttt{amr.n\_cell} & \texttt{256 x 128 x 128} & \texttt{64 x 64 x 512} \\
\quad \texttt{amr.max\_level} & 0 & 0\\
\textbf{Decomposition \& Tiling} & & \\
\quad \texttt{particles.tile\_size} & \texttt{8 x 8 x 8} & \texttt{8 x 8 x 64}  \\
% \quad \textit{Tile Design Note} & \textit{One tile per core target} & \textit{One tile per core target} \\
\textbf{Boundary Conditions} & & \\
\quad Fields & Periodic (all axes) & Periodic (x,y), PEC/PML (z) \\
\quad Particles & Periodic (all axes) & Periodic (x,y), Absorbing/Thermalizing (z) \\
\textbf{Numerics} & & \\
\quad \texttt{warpx.cfl} & 1.0 & 1.0 \\
\quad \texttt{warpx.do\_moving\_window} & 0 (No) & 1 (Yes, along z) \\
\quad \texttt{algo.particle\_shape} & 1 (CIC) and 3 (QSP) & 1 (CIC) \\
\textbf{Plasma Parameters} & & \\
\quad Species & Electrons & Electrons \\
\quad Initial Density & $1 \times 10^{25} \, \text{m}^{-3}$ (uniform) & $2 \times 10^{23} \, \text{m}^{-3}$ (initial background) \\
\quad Momentum & Maxwellian ($u_{\text{th}} = 0.01c$) & Initially at rest, then accelerated \\
\quad Density Dynamics & Homogeneous & Highly non-uniform (bubble, bunches) \\
\textbf{Particle Density Scan (PPC)} & & \\
\quad \texttt{num\_particles\_per\_cell\_each\_dim} & [1,1,1], [2,2,2], [4,4,4], [8,4,4] & [1,1,1], [2,2,2], [4,4,4], [8,4,4] \\
\quad Resulting Average PPC & 1, 8, 64, 128 & 1, 8, 64, 128 \\
\textbf{Laser Parameters} & & \\
\quad Laser present & No & Yes (e.g., Gaussian, $\lambda=0.8\mu m$, $a_0 \sim 1-10$) \\
\quad Continuous injection & None & Yes \\
\textbf{Simulation Length} & & \\
\quad \texttt{max\_step} & 100 (for kernel benchmarks) & 20 (for capturing dynamics) \\
\midrule
\textbf{MatrixPIC Adaptive Sorting Policy} & \multicolumn{2}{c}{\textit{(Applies to FullOpt configurations)}} \\
\quad \texttt{warpx.sort\_interval} & \multicolumn{2}{c}{50} \\
\quad \texttt{warpx.min\_sort\_interval} & \multicolumn{2}{c}{10} \\
\quad \texttt{warpx.sort\_trigger\_rebuild\_count} & \multicolumn{2}{c}{100} \\
\quad \texttt{warpx.sort\_trigger\_empty\_ratio} & \multicolumn{2}{c}{0.15} \\
\quad \texttt{warpx.sort\_trigger\_full\_ratio} & \multicolumn{2}{c}{0.85} \\
\quad \texttt{warpx.sort\_trigger\_perf\_enable} & \multicolumn{2}{c}{1 (Yes)} \\
\quad \texttt{warpx.sort\_trigger\_perf\_degrad} & \multicolumn{2}{c}{0.80} \\
\bottomrule
\end{tabular}%
}
\end{table*}

\twocolumn

\section{Generalizing the MatrixPIC Framework}
\label{appen:generalizing}

The optimization strategies developed in this paper, while demonstrated on the Particle-in-Cell (PIC) current deposition kernel, are not limited to this specific application. They are applicable to a broader class of problems that share a fundamental computational pattern. This appendix details this pattern and illustrates its presence in other critical scientific simulation methods, thereby establishing the generality of the \textbf{MatrixPIC} framework.

\subsection{The Abstract Scatter-Add Computational Pattern}

At its core, the performance bottleneck addressed in this work stems from a fundamental computational pattern: \textit{accumulating a large number of sparse, localized updates onto a regular, dense grid}. This pattern, common in particle-mesh methods, can be deconstructed into three key elements:
\begin{itemize}
    \item \textbf{Source:} A large set of discrete, unordered "particle" entities, each carrying one or more physical quantities (e.g., charge, mass, velocity) and possessing coordinates in a continuous space.
    \item \textbf{Target:} A regular, structured, and dense background grid that represents a field quantity (e.g., density, potential).
    \item \textbf{Operation:} A "scatter-add" or "deposition" process. For each particle, its influence on neighboring grid nodes is computed based on its spatial position, and these contributions are then accumulated onto the corresponding nodes. This process is typically governed by a shape function or kernel function that defines the weighting scheme.
\end{itemize}
The primary performance challenges of this pattern are poor data locality due to the random access of particle data, write contention on the grid, and low arithmetic intensity on traditional architectures.

\subsection{Isomorphic Problems in Scientific Computing}

The following subsections demonstrate how this abstract pattern manifests in three distinct, high-impact scientific domains, making them all suitable candidates for the optimizations presented in \textbf{MatrixPIC}.

\subsubsection{Plasma Physics: PIC Current Deposition}

As the central topic of this paper, the current deposition step in PIC simulations is a canonical example of the scatter-add pattern.
\begin{itemize}
    \item \textbf{Source:} A collection of charged macro-particles, each with a specific charge, mass, and velocity.
    \item \textbf{Target:} A 3D Cartesian grid representing the current density $\mathbf{J}$.
    \item \textbf{Operation:} Each particle's charge and velocity are "scattered" to the 8 neighboring grid nodes using a shape function (e.g., Cloud-in-Cell, CIC) to compute its contribution to the total current density, which is then used to solve Maxwell's equations.
\end{itemize}

\subsubsection{N-body Simulations: PM Mass Deposition}

In cosmological or astrophysical N-body simulations, the Particle-Mesh (PM) method is used to efficiently calculate long-range gravitational forces. Its mass deposition step is algorithmically isomorphic to PIC's current deposition.
\begin{itemize}
    \item \textbf{Source:} A set of galaxies or dark matter particles, each with a specific mass.
    \item \textbf{Target:} A 3D Cartesian grid representing the mass density $\rho$ throughout the simulated cosmic space.
    \item \textbf{Operation:} Each particle's mass is "deposited" onto its 8 neighboring grid nodes using a shape function (often CIC). The resulting mass density on the grid is then used to solve Poisson's equation for the gravitational potential.
\end{itemize}

\subsubsection{Molecular Dynamics: PME Charge Assignment}

In Molecular Dynamics (MD), the Particle-Mesh-Ewald (PME) method is the state-of-the-art technique for calculating long-range electrostatic forces. The first step of this method, known as charge assignment, maps directly onto our abstract pattern.
\begin{itemize}
    \item \textbf{Source:} A collection of atoms, each carrying a fixed partial charge.
    \item \textbf{Target:} A 3D Cartesian grid representing the charge density.
    \item \textbf{Operation:} Each atom's partial charge is distributed or "assigned" to the nearby grid nodes using a high-order shape function (typically a B-spline, which is analogous to the higher-order shape functions in PIC). This grid-based charge density is then used to solve Poisson's equation in Fourier space, enabling an efficient calculation of the long-range forces.
\end{itemize}

In all three cases, despite the different physics being modeled, the underlying computational structure is identical. Consequently, the core performance challenges are shared, and the solutions developed in \textbf{MatrixPIC}—the GPMA sorter to enhance locality, the \texttt{rhocell} concept to mitigate write conflicts, and the MPU mapping to increase computational throughput—are directly applicable to accelerating these other critical scientific applications.

%%%%%%%%%%%%%%%%%%%%%%%%%%%%%%%%%%%%%%%%%%%%%%%%%%%%
% Artifact Appendix for EuroSys'26 AE: MatrixPIC
%
% This document has a maximum length of 2 pages.
%%%%%%%%%%%%%%%%%%%%%%%%%%%%%%%%%%%%%%%%%%%%%%%%%%%%
% \newpage
\section{Artifact Appendix}

\subsection{Abstract}
This artifact provides the implementation of MatrixPIC, an optimized PIC simulation framework built upon the open-source WarpX (v24.07) codebase. The artifact contains the source code modifications that efficiently adapt the computationally-intensive current deposition kernel to emerging CPU architectures equipped with MPUs. It also includes experiment scripts, and per-experiment step-by-step instructions and example logs to facilitate the verification of the paper’s claims.
% and a detailed screen recording for each experiment to facilitate the verification of the paper's claims.

%%%%%%%%%%%%%%%%%%%%%%%%%%%%%%%%%%%%%%%%%%%%%%%%%%%%%%%%%%%%%%%%%%%%%
\subsection{Description \& Requirements}

\subsubsection{How to access}
The artifact, including source code, experiment scripts, and final processed data, is permanently archived on Zenodo at \url{https://doi.org/10.5281/zenodo.17051462}. The development repository is also available for browsing on GitHub at \url{https://github.com/sherry-roar/MatrixPIC_AD.git}. The provided source code consists of the specific files modified within WarpX and AMReX, along with instructions in the \texttt{README.md} file on how to integrate them into the official v24.07 releases of the respective projects.

\subsubsection{Hardware dependencies}
The core contribution of this work is an optimization targeting a specific, next-generation HPC platform, referred to as the "LS pilot system." This system is equipped with LX2 CPUs, which feature the MPUs required to run our optimized code. A detailed description of the hardware is available in Section 5.1 of the main paper.

\textbf{IMPORTANT NOTE:} Due to the highly specific and restricted-access of this hardware, we do not expect the artifact evaluators to be able to run the code. Instead, this artifact is designed to be evaluated by inspecting the source code, analyzing the provided processed data files, and following the detailed step-by-step instructions and example logs that demonstrate the setup, execution, and result collection process on the target hardware.
% and following the detailed screen recordings that demonstrate the setup, execution, and result collection process on the target hardware.

\subsubsection{Software dependencies}
The target platform utilizes a customized software stack, including:
\begin{itemize}
    \item A customized Linux-based operating system.
    \item An LLVM-based Clang/Flang compiler toolchain with architecture-specific auto-vectorization capabilities for the LX2 CPU.
    \item A tuned OpenMPI implementation.
    \item WarpX v24.07 and AMReX v24.07.
\end{itemize}
The \texttt{README.md} file document the environment in detail.

%%%%%%%%%%%%%%%%%%%%%%%%%%%%%%%%%%%%%%%%%%%%%%%%%%%%%%%%%%%%%%%%%%%%%
\subsection{Set-up}
As evaluators cannot directly run the artifact, this section describes the setup process. The \texttt{README.md} file provides a detailed, command-by-command guide for the entire configuration process. Step-by-step instructions and example command logs demonstrate the concrete compilation and execution process for our proposed method on the LX2 CPU platform, mirroring the steps outlined in the \texttt{README.md}.
% The accompanying screen recordings demonstrate the concrete compilation and execution process for our proposed method on the LX2 CPU platform, mirroring the steps outlined in the README. 
The setup for the baseline and GPU-based experiments follows the official WarpX documentation and is therefore omitted here for brevity.

%%%%%%%%%%%%%%%%%%%%%%%%%%%%%%%%%%%%%%%%%%%%%%%%%%%%%%%%%%%%%%%%%%%%%
\subsection{Evaluation workflow}

\subsubsection{Major Claims}
Our paper makes the following major claims, which can be verified using the provided artifact materials:
\begin{itemize}
    \item \textbf{(C1):} MatrixPIC provides significant end-to-end performance speedups over the baseline WarpX for both uniform plasma and LWFA workloads.
    \item \textbf{(C2):} The key components of the MatrixPIC framework contribute measurably to overall performance improvement.
    \item \textbf{(C3):} The incremental sorting algorithm provides a standalone performance benefit, and the complete MatrixPIC framework is faster than a highly-optimized, hand-tuned VPU-based implementation for the 1st-order deposition kernel.
    \item \textbf{(C4):} The performance advantage of MatrixPIC over the best VPU implementation is more substantial for the computationally intensive 3rd-order (QSP) deposition scheme.
    \item \textbf{(C5):} MatrixPIC achieves a higher percentage of theoretical peak hardware performance on an MPU-enabled CPU than the baseline CUDA kernel does on a datacenter GPU.
\end{itemize}
\subsubsection{Experiments}

The following experiments correspond to the major claims. Each experiment’s execution is documented in the \texttt{README.md} and is reproducible using the provided scripts; the resulting processed data is available for inspection in the corresponding spreadsheets.
% Each experiment's execution is captured in the provided screen recordings, and the resulting processed data is available for inspection in the corresponding spreadsheets.

\textbf{IMPORTANT NOTE:}
\begin{itemize}
    \item For brevity, all demonstrations are performed using a particle-per-cell (PPC) density of 1 (\texttt{1x1x1}).
    % To reduce video length, all demonstrations are performed using a particle-per-cell (PPC) density of 1 (\texttt{1x1x1}). 
    The spreadsheets contain the full data across all tested densities.
    \item The execution of the standard WarpX baseline and the GPU baseline is not shown, as their setup follows the official WarpX documentation.
    \item For confidentiality reasons, the previously provided screen recordings have been removed. The scripts, example logs, and spreadsheets are sufficient to verify the paper’s claims.
    % \item All videos are available at the following google drive link: \url{https://shorturl.at/glhqC}
\end{itemize}

\textit{\textbf{Confidentiality Note:} To protect proprietary information about the pre-release hardware and its software environment, the screen recordings have been removed from the artifact package. Please rely on the provided source code, scripts, raw/processed logs, and spreadsheets for verification. Should you encounter any issues with the remaining materials during the evaluation period, please contact the authors.}

% \textit{\textbf{Confidentiality and Video Access Note:} Please be advised that the screen recordings linked in this document contain proprietary information related to the pre-release hardware platform and its software environment. These videos are made available exclusively for the purpose of artifact evaluation. The access link will remain active throughout the entire artifact evaluation period and will be disabled upon the completion of the review process. We kindly request that the video content not be shared or distributed. Should you encounter any access issues during the evaluation period, please contact the authors immediately.}

\textit{\textbf{Experiment (E1)}: Overall Performance in Uniform and LWFA Scenarios:} This experiment demonstrates the end-to-end workflow for running MatrixPIC and post-processing the results to calculate performance metrics.
% (Videos: \texttt{6-1cal\_matrix}, \texttt{6-1order1marixPIC-LWFA}, \texttt{6-1order1matrixPIC})

\textit{[Preparation]} The provided step-by-step instructions (in \texttt{README.md}) show how to modify the input files (\texttt{input\_files/inputs3d-l} or \texttt{input\_files/LWFA}) to set parameters like particle density and shape factor. It also demonstrates how to select the MatrixPIC kernel by uncommenting the \texttt{doDepositionShapeN\_3d\_sme<1>()} function in the source code.

\textit{[Execution]} Execution uses the main run script (\texttt{build\_release\_and\_run.sh}) and the Python post-processing script (\texttt{calculate\_particle\_avg\_metric.py}), as described in the \texttt{README.md}.

\textit{[Results]} The final performance data is available in the spreadsheet \texttt{eurosys\_outputs/sec6.1.xlsx}, corresponding to Figures 8 and 9. The artifact includes the raw logs and documents, in the \texttt{README.md}, how the post-processing script is used to derive the final metrics found in the spreadsheet.
% The video demonstrates the execution that generates the raw logs and how the post-processing script is used to derive the final metrics found in the spreadsheet.

\textit{\textbf{Experiment (E2)}: Ablation Study:} This experiment isolates the performance impact of each component of the MatrixPIC framework.
% (Videos: \texttt{6-2hybrid-nosort}, \texttt{6-2pure-sme}, \texttt{6-2sort-real})

\textit{[Preparation]} The \texttt{README.md} describes how to activate different kernels (Matrix-only, Hybrid-noSort, Hybrid-GlobalSort) in the source code. For configurations without global sorting, the video shows how the feature is disabled by setting \texttt{min\_sort\_interval} to a large value. Other parameters are identical to the 1st-order FullOpt experiment in (E1).

\textit{[Execution]} Execution follows the same procedure as in Experiment (E1).

\textit{[Results]} Final performance data is provided in the spreadsheet \texttt{eurosys\_outputs/sec6.2.xlsx}, corresponding to Figure 10. A sample run for one of the ablation configurations is documented in the \texttt{README.md}.

\textit{\textbf{Experiment (E3)}: Comparison with Optimized VPU Kernels (1st-Order CIC):} This experiment compares MatrixPIC against optimized VPU baselines and demonstrates the CPU cycle-to-time conversion process.
% (Videos: \texttt{6-3order1org-gpma}, \texttt{6-3order1rhocell}, \texttt{6-3order1rhocell-gpma}, \texttt{6-3-transfertcycle2time})

\textit{[Preparation]} Preparation is similar to Experiment (E2), involving the selection of different VPU-based kernels in the source code.

\textit{[Execution]} Execution follows the procedure in (E1). The \texttt{README.md} documents how to use the \texttt{utils/transfer\_cycles2times.sh} script.

\textit{[Results]} The final timing data is available in the spreadsheet \texttt{eurosys\_outputs/sec6.3.xlsx}, corresponding to Table 1. The provided utility script and the \texttt{README.md} show how the raw output (in CPU cycles) is converted into seconds.

\textit{\textbf{Experiment (E4)}: Higher-Order QSP Kernel Performance (3rd-Order QSP):} This experiment demonstrates the 3rd-order QSP performance for a key baseline (Baseline+IncrSort). The runs for MatrixPIC and the optimized VPU kernel are shown in (E5).
% (Video: \texttt{6-3order3org-gpma})

\textit{[Preparation]} Preparation is identical to Experiment (E1), except the \texttt{shape\_factor} is set to 3 in the input file to enable the 3rd-order scheme.

\textit{[Execution]} The simulation is run with a single MPI process and a single OpenMP thread to measure single-core performance.

\textit{[Results]} The final data is in the spreadsheet \texttt{eurosys\_outputs/sec6.3.xlsx}, corresponding to the relevant rows in Table 2.

\textit{\textbf{Experiment (E5)}: Cross-Platform Peak Performance Efficiency:} This experiment demonstrates the 3rd-order QSP performance for the optimized MatrixPIC and VPU kernels, and shows how peak performance efficiency is calculated.
% (Videos: \texttt{6-4order3MatrixPIC}, \texttt{6-4order3rhocell-sort-vpu})

\textit{[Preparation]} Preparation is the same as in Experiment (E4).

\textit{[Execution]} Execution follows the multi-threaded procedure from Experiment (E1).

\textit{[Results]} The final performance data corresponds to Tables 2 and 3. The \texttt{README.md} explains where to find the \texttt{Peak Performance} metric in the output logs. For cross-platform validation, the raw output log from the GPU baseline run is also provided in \texttt{eurosys\_outputs/sec6.4-gpu.out}.


\end{document}